\documentclass[lettersize,journal]{IEEEtran}
\usepackage{amsmath,amsfonts}
\usepackage{algorithmic}
\usepackage{array}

\usepackage{siunitx}
\usepackage{xspace}
\usepackage{textcomp}
\usepackage{stfloats}
\usepackage{url}
\usepackage{verbatim}
\usepackage{graphicx}
\usepackage{color}
\usepackage{soul}
\usepackage{flushend}
\usepackage{balance}
\usepackage{subfigure}

\newcommand{\sys}{\textit{TimeSense}} 
\newcommand{\sysx}{\textit{TimeSense}}

\begin{document}
\title{\sysx{}: Multi-Person Device-free Indoor Localization via RTT}




\author{
    Mohamed Mohsen,
    Hamada~Rizk,~\IEEEmembership{Senior Member,~IEEE}, and
    Hirozumi~Yamaguchi,~\IEEEmembership{Senior Member,~IEEE}, and
    Moustafa~Youssef,~\IEEEmembership{Fellow Member,~IEEE}
         \thanks{This work was partially supported by the Grant-in-Aid for Scientific Research (C) (Grant number 22K12011) from JSPS (Japan Society for the Promotion of Science).}
          \thanks{M. M. is with Benha University, Cairo 11637, Egypt (e-mail:mohamed.hekal@feng.bu.edu.eg). }
        \thanks{H. R. (Corresponding Author) is with Tanta University, Tanta 31733, Egypt, and Osaka University, 565-0871 Japan (e-mail: hamada\_rizk@f-eng.tanta.edu.eg). }
\thanks{H. Y.  is with Osaka University, 565-0871 Japan (e-mail: t-h-yamagu@ist.osaka-u.ac.jp).}
\thanks{M. Y. is with 
 is with AUC, Cairo, Egypt (e-mail: moustafa-youssef@aucegypt.edu).}
}





\maketitle

\begin{abstract}
Locating the persons moving through an environment without the necessity of them being equipped with special devices has become vital for many applications including security, IoT, healthcare, etc. 
Existing device-free indoor localization systems commonly rely on the utilization of Received Signal Strength Indicator (RSSI) and WiFi Channel State Information (CSI) techniques. However, the accuracy of RSSI is adversely affected by environmental factors like multi-path interference and fading. Additionally, the lack of standardization in CSI necessitates the use of specialized hardware and software.
In this paper, we present \sys{}, a deep learning-based multi-person device-free indoor localization system that addresses these challenges. \sys{} leverages Time of Flight information acquired by the fine-time measurement protocol of  IEEE 802.11-2016 standard.
Specifically, the measured round trip time between the transmitter and receiver is influenced by the dynamic changes in the environment induced by human presence. \sys{} effectively detects this anomalous behavior using a stacked denoising auto-encoder model, thereby estimating the user's location. The system incorporates a probabilistic approach on top of the deep learning model to ensure seamless tracking of the users.
The evaluation of \sys{} in two realistic environments demonstrates its efficacy, achieving a median localization accuracy of 1.57 and 2.65 meters. This surpasses the performance of state-of-the-art techniques by 49\% and 103\% in the two testbeds.
\end{abstract} 


\section{Introduction}
Indoor localization has garnered significant attention due to its critical importance across various applications, such as industrial operations, emergency response services, security measures, and logistical management \cite{buyukccorak2014indoor, emam2017adaptive, bahl2000radar,cellindeep2019,wideep2019}. 

There are two primary approaches to localization systems: device-based and device-free solutions.
Device-based localization systems \cite{rizk2022robust,hashem2020deepnar, hashem2020winar,rizk2022vaccinated} require individuals to carry specific hardware, such as smartphones, to facilitate their localization. While GPS is the standard for outdoor localization, its effectiveness indoors is compromised due to signal blockage \cite{aly2017accurate}, leading to the exploration of alternative technologies including Wi-Fi, Bluetooth, Ultra-Wideband (UWB), and cellular networks \cite{zafari2019survey,al2011survey,mautz2009overview,cellindeep2019,monodcell2019,rizk2020ubiquitous,rizk2020omnicells}.

On the other hand, device-free indoor localization becomes essential in scenarios where it is important to track individuals who may not be carrying any devices, such as in elderly monitoring applications. This approach is pivotal in situations requiring localization without dependence on personal devices \cite{mohsen2023locfree}.
Camera-based systems are prevalent in device-free localization but encounter several challenges \cite{krumm2000multi}. These include dependency on line-of-sight, leading to substantial deployment costs to ensure complete coverage, poor performance in low-light or smoky conditions, and potential privacy issues. Consequently, there is a growing interest in leveraging alternative methods, such as Wi-Fi-based technologies that utilize radio waves, to circumvent these challenges.

The received signal strength indicator (RSSI) of Wi-Fi signals is significantly influenced by human-body blockages, creating an opportunity for its utilization in device-free localization \cite{youssef2007challenges,moussa2009smart}. However, indoor Wi-Fi RSSI-based systems encounter several challenges, including signal fading due to obstructions such as walls and objects, signal fluctuations caused by multi-path fading and radio interference, and variations in access points' transmission power to accommodate varying traffic demands. These challenges result in a degradation of the performance of Wi-Fi RSSI-based systems \cite{feng2022analysis}.
Conversely, Wi-Fi channel state information (CSI) has been leveraged by many systems \cite{wang2016lifs,jiang2020towards, venkatnarayan2020leveraging,karanam2019tracking} as it exhibits sensitivity to changes in radio waves and offers valuable indications of human presence. Nonetheless, the lack of standardization for CSI necessitates the use of specialized hardware or software to acquire it, rendering it impractical for numerous applications.

Recently, time-based techniques have shown promising solutions, particularly in device-based settings. These techniques estimate the distance between a mobile device (e.g., smartphone) and access points (APs) by measuring the signal's propagation time and utilizing the known propagation velocity of the signal. Various approaches have been proposed for measuring propagation time, including time of arrival (ToA) \cite{golden2007sensor}, time difference of arrival (TDoA) \cite{chan2006time,mohsen2023passifi}, and round-trip time (RTT) \cite{hashem2020winar}. ToA and TDoA methods necessitate precise time synchronization among all devices, posing a challenge. In contrast, RTT utilizes the difference in recorded times to measure the time required for the signal to travel to a destination node and return, thereby mitigating the synchronization problem.
Unlike RSSI-based techniques, RTT demonstrates greater resilience to the challenges posed by complex indoor environments \cite{feng2022analysis}. Furthermore, the fine time measurement (FTM) protocol, which enables the measurement of RTT between mobile phones and APs, has recently been standardized by IEEE 802.11mc. This protocol has gained increasing support from commercial APs and consumers' mobile phones, thereby making time-based techniques a promising solution for enabling practical indoor localization \cite{hashem2020deepnar,hashem2020winar,rizk2015hybrid}.

In this paper, we present \sys{}, a novel device-free  multi-person indoor localization system that effectively utilizes the advantages of RTT  to offer a practical and robust solution.
Specifically, it leverages the changes in RTT measurements caused by the presence of one or more humans, as received by multiple receivers in a device-free setting. However, \sys{} has to face the challenge of the inherent error proneness of RTT measurements, due to latency in non-line-of-sight transmissions, which results in signals taking longer, indirect paths and leading to the overestimation of travel distances in device-free localization \cite{ibrahim2018verification}. This issue becomes even more pronounced in scenarios such as multi-user localization, which either requires complex models that are not easily scalable with area expansion or are sensitive to environmental variations. To address these challenges, \sys{} constructs multiple shallow networks, each tailored to a specific reference point in the environment where RTT measurements were collected.
 Specifically, the system leverages the power of shallow denoising autoencoders to enhance the system's robustness against noisy and distorted measurements.
 This design promotes scalability and facilitates easy integration of new models as the area expands, without needing to retrain existing models. These shallow autoencoders are faster and more scalable compared to a single, deep model that requires extensive complexity and volume of training data.
Through careful training on different types of noisy signals, the system ensures generalization across various environmental conditions and minimizes the impact of signal distortions. Lastly, \sys{} employs diverse regularization techniques to prevent overfitting during the training process, thereby ensuring the model's robustness and enhancing its generalization capability.

For system evaluation, we deployed \sys{} using different APs, and android devices on two different real-world environments, a large environment with an area of $340 m^2$ and a small one of $48 m^2$.
Our results show that \sys{} achieved a median localization error of $1.57$ $meter (m)$ and $2.65m$ for the two environments, respectively. These results reveal an improvement over the traditional RSSI accuracy by at least 49\%.

The contribution of this paper is fourfold. First, we present a practical deep-learning-based device-free indoor localization system leveraging the standardization of the FTM protocol on consumer devices. Second, we increase the system's robustness to any abrupt noise by leveraging the power of denoising auto-encoders. Third, we enable the localization of multiple persons in the continuous space using a probabilistic framework. 
Finally, we experimentally validate the system's ability to locate multiple persons in the realistic cluttered testbed.

The rest of the paper is organized as follows: Section \ref{sec:Related Work} discusses related work. Section \ref{Background and Discussion} gives a brief explanation of the Wi-Fi FTM protocol. The basic idea behind \sys{} is discussed in Section \ref{Device-free Pedestrian Indoor Localization Basic Idea}. In Section \ref{Sys_overview}, an overview of the system’s building blocks and the mathematical model are introduced.
In Section \ref{WiDAD System} the modules of the system are discussed in detail. We evaluate the system performance in Section \ref{Evaluation}. Finally, Section \ref{sec:Conclusion} concludes the paper. 

\section{Related Work} \label{sec:Related Work}
Various technologies have been employed by diverse methodologies to deliver localization solutions. The widespread presence of WLANs in most indoor environments has captured the interest of researchers, prompting the development of solutions that leverage WiFi network infrastructure to offer localization services with a high level of accuracy. This section continues by reviewing prior studies in both device-based and device-free localization systems.

\subsection{Device-based Indoor Localization Systems}
 WiFi is among the most leveraged technologies in indoor localization solutions due to its ubiquitous availability. In order to fulfill the location determination task, these systems rely on the RSSI and the RTT techniques, often in conjunction with fingerprinting methods.
In the RSSI-based fingerprinting approaches \cite{youssef2005horus, he2015wi, wideep2019, rizk2023indoor, rizk2024adaptability, elkholy2023virtual}, the RSSI measurements are collected from the surrounding access points (APs) at a number of predefined reference points covering the area of interest. Then, these measurements are utilized to build a radio map which is leveraged to construct a localization model that is capable of estimating the user's location given the received RSSI readings. 
A probabilistic framework is utilized in the Horus system \cite{youssef2005horus} to fulfill the localization task. This is achieved by generating probability distributions of the RSSI values from different APs at each location. Then, determine the user's location by determining the highest likelihood of the received RSSI values when compared to the pre-collected RSSI data. The system in \cite{jang2018indoor} trains a convolutional neural network (CNN) using the collected RSSI fingerprints. This CNN-based system shows a higher localization accuracy than the probabilistic model, revealing the potential of deep-learning models in indoor localization solutions. Additionally, WiDeep \cite{wideep2019} leverages the constructed RSSI radio map to train multiple stacked denoising autoencoders for latent feature extraction. Additionally, it employs a probabilistic model for more accurate tracking in the continuous space.

Due to its resilience and robustness in indoor environments, the round-trip-time (RTT) technique gained more traction in recent years. 
WiNar \cite{hashem2020winar} utilizes RTT fingerprints to construct a probabilistic model based on Bayesian inference. This model estimates the probability of the user's presence at predefined reference points, providing valuable information for localization.
DeepNar \cite{hashem2020deepnar} leverages the collected RTT data during the offline phase to train a multi-layer deep-learning model acting as a multi-class classifier. During the online phase, the user's device captures RTT measurements from nearby access points and feeds them to the trained model, which generates the probability of the user's existence at the reference points.
On the other hand, RRLoc \cite{rizk2022robust} produces an enhanced performance compared to previous RTT- and RSSI-based systems. It employs a hybrid technique that combines both RSSI- and RTT-based measurements, integrating them through a DeepCCA network to extract high-level features. These features are then utilized to train a deep classification model for accurate localization.
Finally, MagTT \cite{Farah2022MagttLoc}, incorporates magnetic field measurements with RTT to achieve submeter-level accuracy. Leveraging the power of a CNN-LSTM architecture, MagTT demonstrates the potential of combining different sensor data for enhanced indoor localization accuracy.

While RSSI-based systems have the advantage of not requiring specific hardware and using signal strength for localization, they are susceptible to obstacles, interference, and multipath effects. In contrast, RTT provides more accurate distance measurements by directly capturing signal propagation time delay. The standardization of RTT by IEEE 802.11mc has made it widely available in commercial-off-the-shelf (COTS) devices like smartphones and access points. 

\textit{The standardization of the RTT-based technique in the WiFi technology and the robustness of the time-based techniques gave the localization systems the ability to present an enhanced performance with fine-grained accuracy. Motivated by this, our work focuses on leveraging RTT measurements as features to localize multiple persons within an indoor environment in a device-free fashion.}

\subsection{Device-free Indoor Localization Systems}
In device-free systems, the localization of persons is achieved without the necessity of requiring them to carry a dedicated device. Different technologies were leveraged by several proposed systems including ultra-wideband (UWB) \cite{ram2007human}, computer vision \cite{krumm2000multi}, and LiDAR \cite{ohno2023privacy, liphi}, which necessitate specialized hardware for their operation.

Computer vision-based systems heavily rely on camera sensors \cite{ram2007human}, which have limitations in non-line-of-sight (NLoS) conditions, leading to high deployment costs when covering large areas. Moreover, regular cameras fail to function effectively in low light or the presence of smoke, and in certain scenarios, they may raise privacy violation concerns related to the persons in some applications \cite{ohno2023privacy, liphi}.

On the contrary, Wi-Fi-based systems \cite{youssef2007challenges, moussa2009smart, 6199865, wang2016lifs} are built upon the existing infrastructure of WLAN networks without requiring additional hardware. The pioneering Wi-Fi-based device-free indoor localization system proposed in \cite{youssef2007challenges} utilizes Wi-Fi RSSI to construct a radio map within the area of interest. When disturbances occur due to the movement of individuals, a probabilistic model compares the collected RSSI data to the pre-collected data during the offline phase to perform accurate localization.
RASID \cite{6199865} employs a semi-supervised statistical technique to extract statistical features, such as measures of central tendency (e.g., mean) and measures of dispersion or variation (e.g., variance), from the input data. These measures/features are used to construct a density function that sets a threshold value for detecting anomalous behaviors and locating their source of anomaly. 
Wi-Fi Channel State Information (CSI) \cite{wang2016lifs}, which is extracted information from radio waves such as amplitude and phase difference. CSI responds to disturbances in the radio waves; therefore, it has been researched for human-activity recognition \cite{chen2018wifi},
In \cite{wang2016lifs}, the system leverages CSI to formulate and solve power fading model equations for all the links to achieve the localization task.

\textit{On the other hand, \sys{} is based on RTT which is more robust than RSSI. In addition, RTT is feasible as it is standardized by IEEE 802.11mc which makes it widely available in consumer devices. Furthermore, \sys{} leverages a deep-learning denoising auto-encoder for more robustness against any abrupt noise.}

\begin{figure}[t]
 \centering
 \begin{minipage}[b]{0.94\linewidth}
    \includegraphics[width=\linewidth,height=6.0cm,]{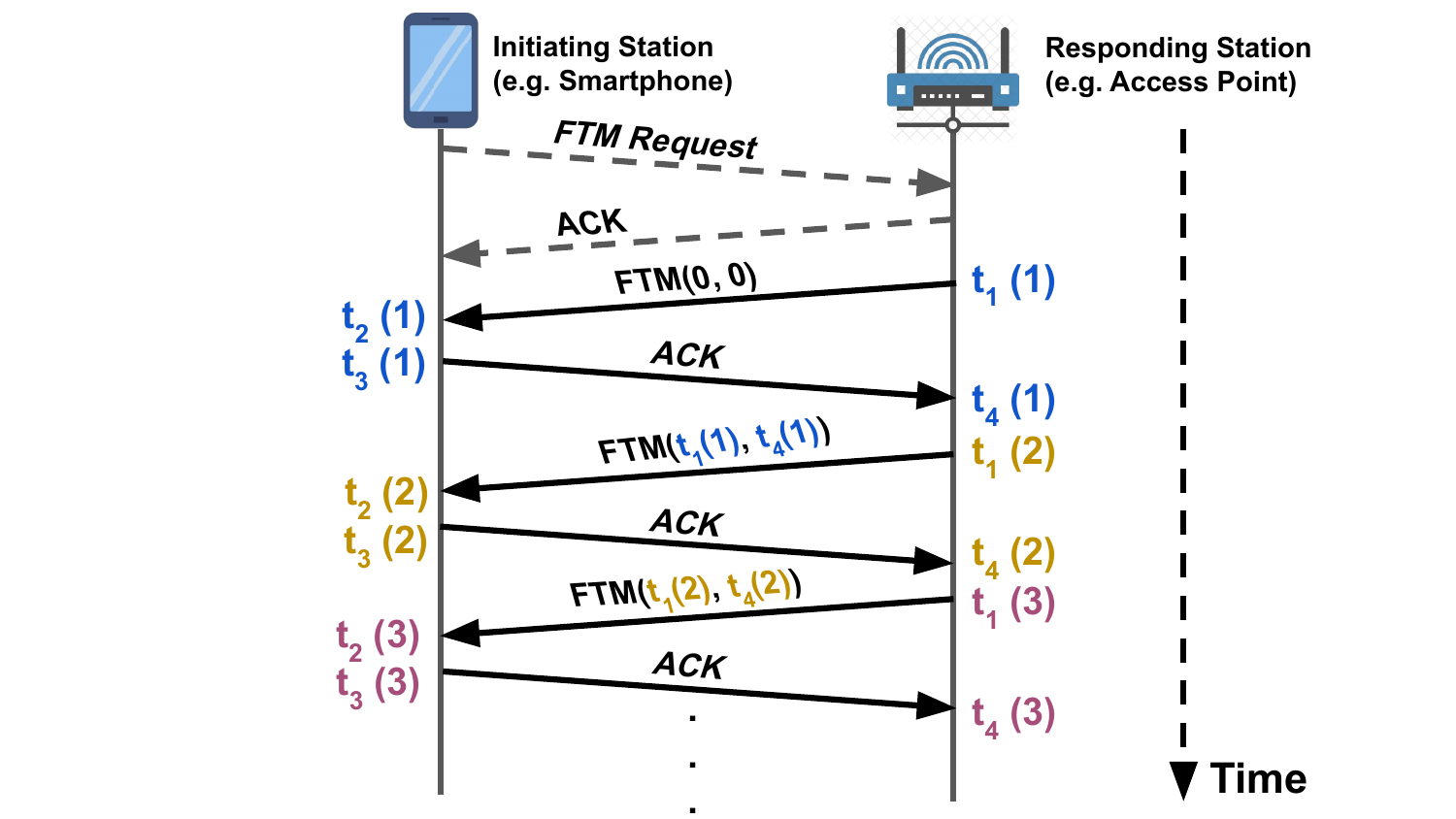}
    \caption{The basic RTT distance estimation procedure in the IEEE 802.11mc standard.}
    \label{fig:FTM_proc}  
 \end{minipage}
\end{figure}

\section{Wi-Fi FTM Protocol Background} \label{Background and Discussion}
In 2016 the IEEE 802.11mc standardized the Fine-Timing Measurement (FTM) protocol that gives a Wi-Fi device the ability to accurately estimate the round trip time (RTT) between it and another Wi-Fi device. Fig.~\ref{fig:FTM_proc} shows the Wi-Fi FTM basic procedure. This process involves two types of devices: the initiating station, usually a smartphone, and the responding station which is usually applied by an access point (AP). The measurement procedure starts with the initiating station sending an FTM request to the responding station to check if it is available and to negotiate some parameters regarding the FTM process such as the number of bursts, the number of measurements per burst, etc. If the responding station is available, it replies with an acknowledgment frame (ACK) back to the initiating station.

After the initiating phase, the responding station sends an FTM frame to the initiating station and records the frame's time of departure (i.e., $t_1$). On receiving the FTM frame, the initiating station records the frame's time of arrival (i.e., $t_2$). After some while, the initiating station replies back to the responding station with an ACK frame and records the frame's time of departure (i.e., $t_3$). Finally, the responding station receives the ACK frame, sent by the initiating station, and records the frame's time of arrival (i.e., $t_4$). The responding station sends the values of $t_1$ and $t_4$ in the next FTM frame in the same burst which gives the initiating station to calculate the round trip time (RTT) as follows:
\begin{equation} \label{RTT_calc}
    \mathrm{RTT}=\left(t_4-t_1\right)-\left(t_3-t_2\right)
\end{equation}
To provide a more accurate RTT estimation, the FTM interchanges can be repeated multiple times in the form of bursts, then the average RTT is calculated as follows:
\begin{equation} \label{RTT_avg_calc}
    AvgRTT= \frac{1}{N}\sum_{i=1}^{N}( \left(t_4(i)-t_1(i)\right) - \left(t_3(i)-t_2(i)\right) ) 
\end{equation}
Where N is the number of exchanged FTM frames in a burst. Then the distance between the initiating and the responding stations can be estimated as follows:
\begin{equation} \label{RTT_distance}
    distance= \frac{AvgRTT}{2} \times \mathrm{c}
\end{equation} Where c is the propagation speed of light in space ($3 \times 10^8$ m/s ).

In order for the initiating station (i.e., smartphone) to locate itself, it performs an RTT ranging process with all the available APs. Collecting enough RTT data, the localization task can be achieved using several algorithms such as multi-lateration \cite{ciurana2007ranging, ibrahim2018verification} or deep learning \cite{hashem2020deepnar, rizk2015hybrid}.

\begin{figure}[t]
 \centering
 \begin{minipage}[b]{0.94\linewidth}
\includegraphics[width=\linewidth]{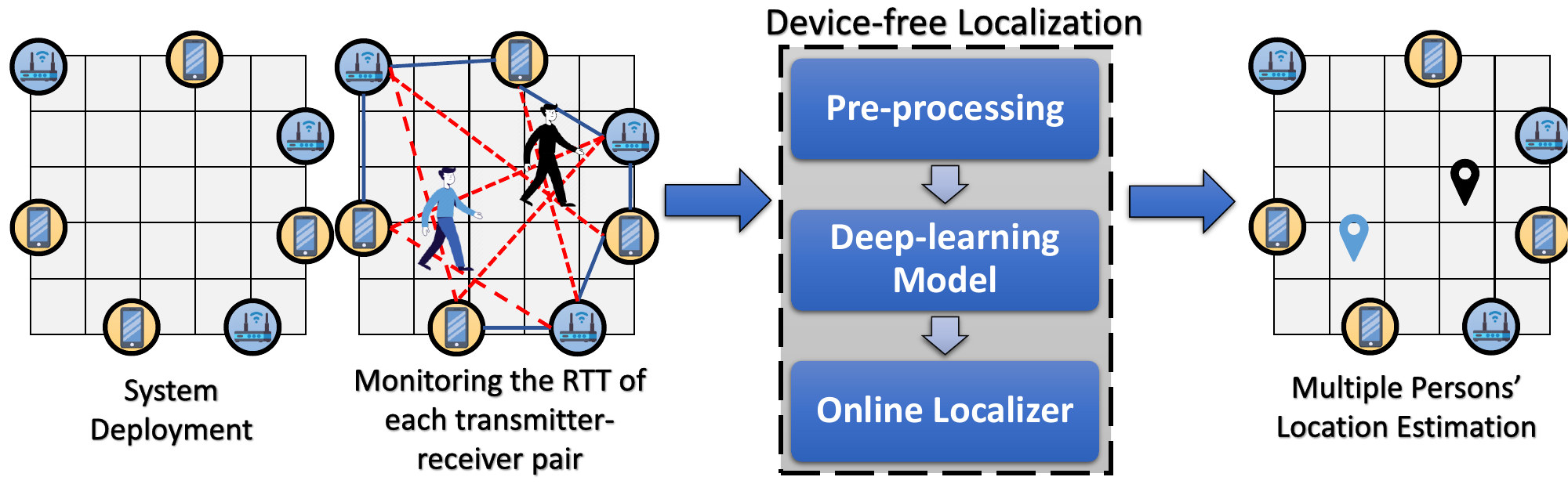}
    \caption{The basic procedure of a device-free indoor localization system. {\color{black}The RTT measurements corresponding to a person's blockage to the LoS between a collection of transmitter-receiver pairs is fed to the model for localization.}}
    \label{fig:Device_free_detection}  
 \end{minipage}
 \vspace{-0.5cm}
\end{figure}

\section{ The Basic Idea } \label{Device-free Pedestrian Indoor Localization Basic Idea}
\begin{figure}[t]
 \centering
 \begin{minipage}[b]{0.94\linewidth}
\includegraphics[width=\linewidth,height=5.0cm,]{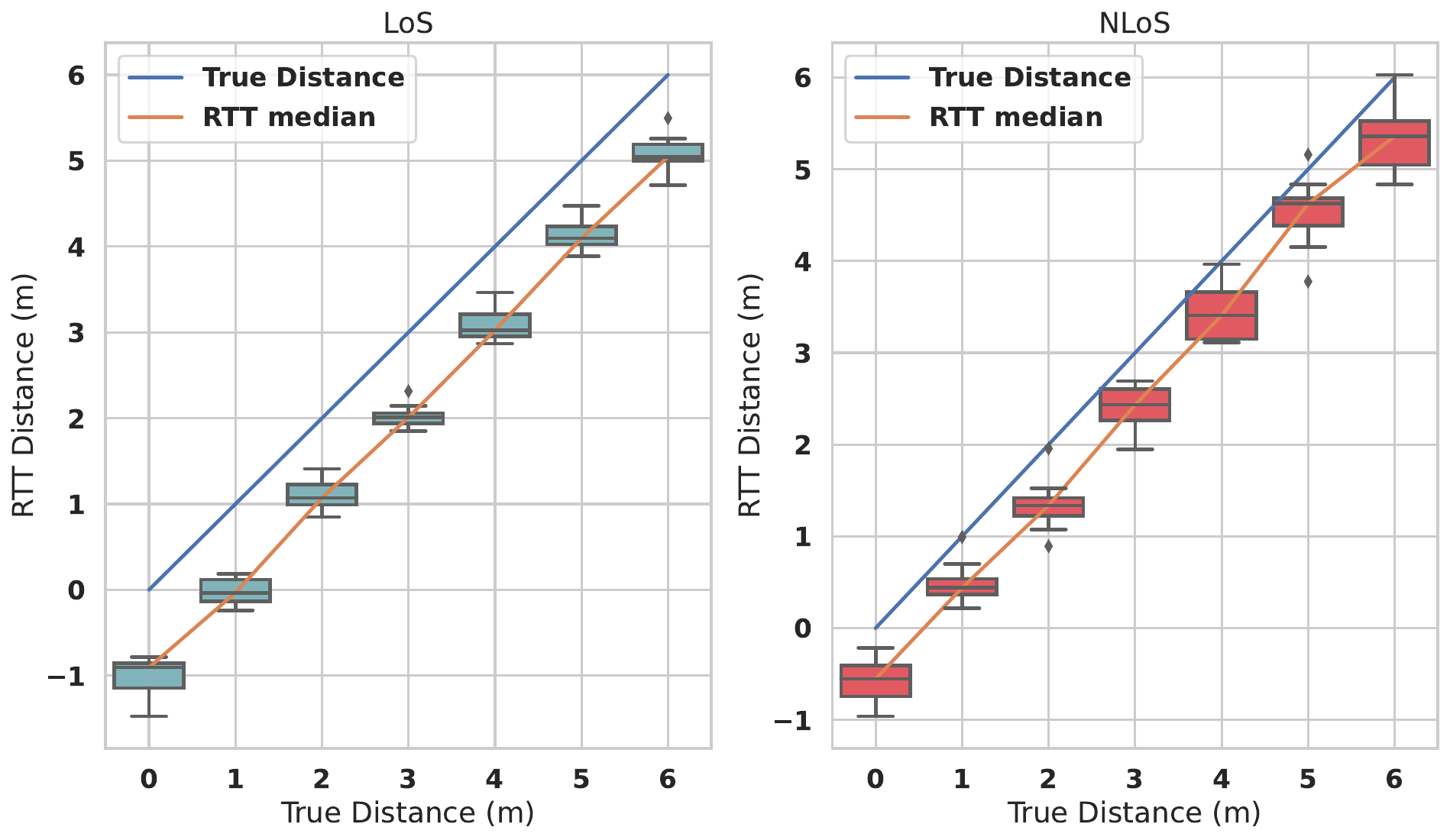}
    \caption{The resulting change in the RTT measurements due to the human body blockage. }
    \label{fig:disturbance}  
 \end{minipage}
  \vspace{-0.5cm}
\end{figure}

The fundamental concept behind \sys{} is to leverage the monitoring of signal propagation time between transmitters (e.g., APs) and receivers (e.g., smartphones) to detect the presence of individuals in the environment. This is accomplished by analyzing whether radio waves have been obstructed (blockage occurred) for each transmitter-receiver pair.
When a person is present, their presence disrupts the direct line-of-sight (LoS) path between the transmitter and receiver, causing the signal to travel a longer distance path, indirect non-line-of-sight (NLoS), causing an increase in the time measurements. This effect is due to blockage caused by the human body, as illustrated in Figure \ref{fig:Device_free_detection}. By examining the combination of all NLoS and LoS paths of the transmitter-receiver pairs, the system can determine the user's location.
However, cluttered environments and the dense presence of crowds can introduce variations in the RTT measurements of transmitter-receiver pairs that traverse the user's location, as illustrated in Figure~\ref{fig:disturbance}. In this vein, RTT measurements are susceptible to errors arising from device offsets, measurement noise, multipath effects, environmental influences, and variances caused by different individuals. To address this challenge, we employ a passive fingerprinting approach by capturing RTT measurements at each reference point. Subsequently, a denoising autoencoder is utilized to learn the signal state specific to each reference point. This enables the system to discern whether the user is present at a given point, thereby facilitating accurate localization in a device-free manner.

\section{System Overview and Mathematical Model} \label{Sys_overview}
\subsection{\sys{} System Overview}
\begin{figure}[t]
 \centering
 \begin{minipage}[b]{0.94\linewidth}
\includegraphics[width=\linewidth, height=6.0cm]{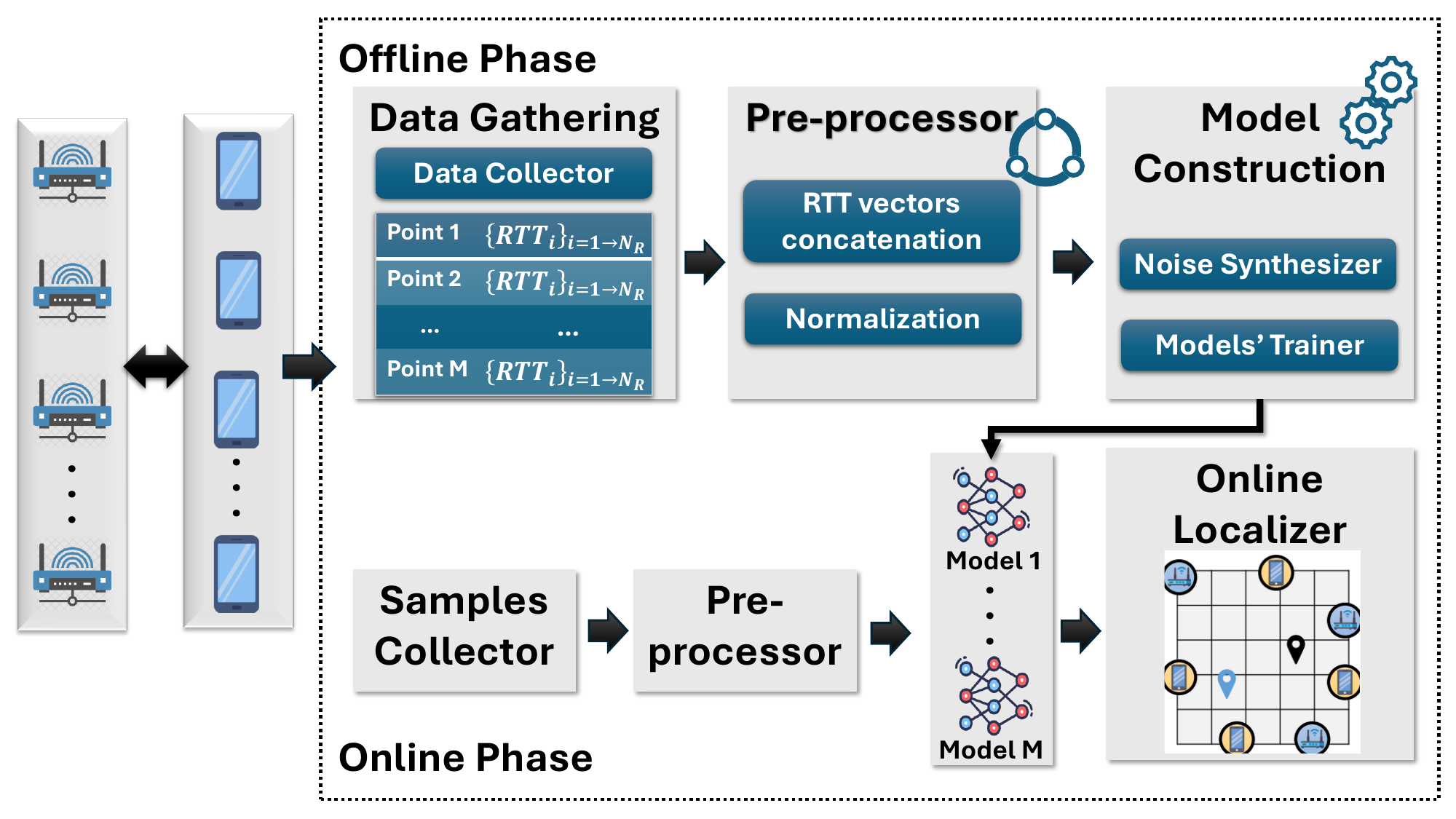}
    \caption{\color{black}\sys{} System Architecture.}
    \label{fig:WiDAD_SysArch}  
 \end{minipage}
\end{figure}

The system architecture of \sys{} is illustrated in Figure \ref{fig:WiDAD_SysArch} and comprises two phases: the offline training phase and the online localization phase.
During the offline training phase, the system constructs $M$ deep neural networks, each corresponding to a specific reference point. This approach enables scalability for larger areas while maintaining smaller and more manageable model sizes. The training data is collected using the \textbf{Data Collector} module, which records the signal states when a person is located at designated reference points within the area of interest. This involves capturing the RTT measurements of each transmitter-receiver pair. The data collection process is repeated for each reference point in the environment. The collected data is then transmitted to the cloud-based \sys{} running service for processing. The \textbf{Pre-processor} module appropriately formats the data to facilitate the training of deep learning models, and the readings are normalized to enhance the training process's efficiency. To introduce noise and prevent overfitting, the preprocessed data is fed to the \textbf{Noise Injector} module, which corrupts the original measurements by injecting artificial noise. This technique helps simulate the presence of noisy measurements and encourages the models to learn the representative features of the signal state. Next, the corrupted and original data for each reference point is passed to the \textbf{Model Construction} module, which creates and trains stacked denoising autoencoders specific to each reference point. 
Finally, all the trained models for the different reference locations are stored for subsequent use during the online localization phase.

In the online localization phase, the system provides a real-time estimation of the persons' unknown locations. The system continually collects and pre-processes the current system state, ensuring proper shaping and normalization of the data. This processed information is then fed into the pre-trained models associated with different reference points. The \textbf{Online Localizer} module employs a probabilistic framework, leveraging the outputs of the various deep learning models, to estimate the persons' locations. This estimation process considers the probabilities generated by each model, resulting in a robust and accurate localization outcome.

\subsection{Mathematical Model}
Assuming a 2D area of interest $\mathbb{L}$ where $N_T$ access points (Transmitters) and $N_R$ Android devices (Receivers) are deployed. The area is divided into a definite number of reference points $M$. The data for each point is collected while keeping a person standing at that point. During the online phase, a pedestrian moves through an unknown location $l_i \in \mathbb{L}$ cutting the line of sight (LoS) between a number of transmitter-receiver pairs causing a disturbance in the system state. Let a vector $s_t$ of $K$ dimensions represent the system state at some time $t$. Each entry in this vector represents the round trip time $RTT_{ij}$ estimated between the transmitter $i$ and the receiver $j$. The problem is: given the system state $s_t = <RTT_{11}, RTT_{12}, ..., RTT_{21}, RTT_{22}, ...., RTT_{N_T, N_R}>$ $\in \mathbb{R}^K$ ($K = N_T \times N_R$), the objective is to detect the abnormality in the state vector entries if existed which indicates that some pedestrian is passing and find the location $l_i$ that maximizes the probability $P(l_{i}\mid s_t)$. In the next section, we discuss in detail the \sys{} system's building blocks and how they are combined together to achieve the localization task.

\section{The \sys{} System} \label{WiDAD System}
\subsection{The Pre-processing Module}
This module plays an essential role in both the offline and online phases of \sys{}, serving multiple functions with a technical focus on data preparation and anomaly handling for deep-learning models. It processes RTT data from all receivers to compile an input vector for the deep-learning model, with each element representing an RTT measurement between a receiver and an access point. This module efficiently handles abrupt changes in network infrastructure or unexpected behavior stemming from configuration discrepancies.  Specifically, it has been observed that not all installed APs are consistently detected in each scan, leading to input vectors of variable lengths. To mitigate this variability, APs undetected in a particular scan are represented with a placeholder RTT value of 
 $0.2\times10^{-3}milliseconds$, equivalent to an estimated distance of 60m, exceeding typical RTT values for APs within the detection range. 
Furthermore, this module addresses anomalies such as the Android API reporting negative distances when a mobile device is in close proximity to an AP. This issue can be traced back to the internal configurations and calibration of WiFi cards, or to multipath compensation algorithms influencing the measurements pre-reception by the device driver. RTT measurements may also exhibit latency anomalies at certain reference points. Traditional multi-lateration approaches often suffer in accuracy due to the presence of negative values or latency \cite{ibrahim2018verification}. Contrarily, \sys{} maintains robust performance in the face of such anomalies, leveraging negative values and delays as distinctive signatures for specific locations within its fingerprinting-based technique.
Finally, the pre-processor normalizes the input vector's features, ensuring they fall within a $[0, 1]$ range. This normalization significantly enhances the optimization process during the model training phase, optimizing \sys{}'s overall efficacy and precision in localization.

\subsection{The Model Construction Module}

The primary function of this module is to develop and train a collection of 
M shallow neural networks, each customized for a unique reference point. Utilizing a straightforward vanilla model for each point, the system skillfully handles various challenges, such as the inherent noise and fluctuations in RTT measurements. To enhance the models' generalization capabilities and ensure their adaptability to scenarios with varying numbers of individuals at neighboring points, the module employs strategies to prevent overfitting the training data. Furthermore, the design of this module supports extendibility, enabling it to seamlessly adjust to expansions in the environmental space, thereby ensuring its scalability and flexibility.

\begin{figure}[t]
 \centering
 \begin{minipage}[b]{0.94\linewidth}
\includegraphics[width=\linewidth, height=4cm]{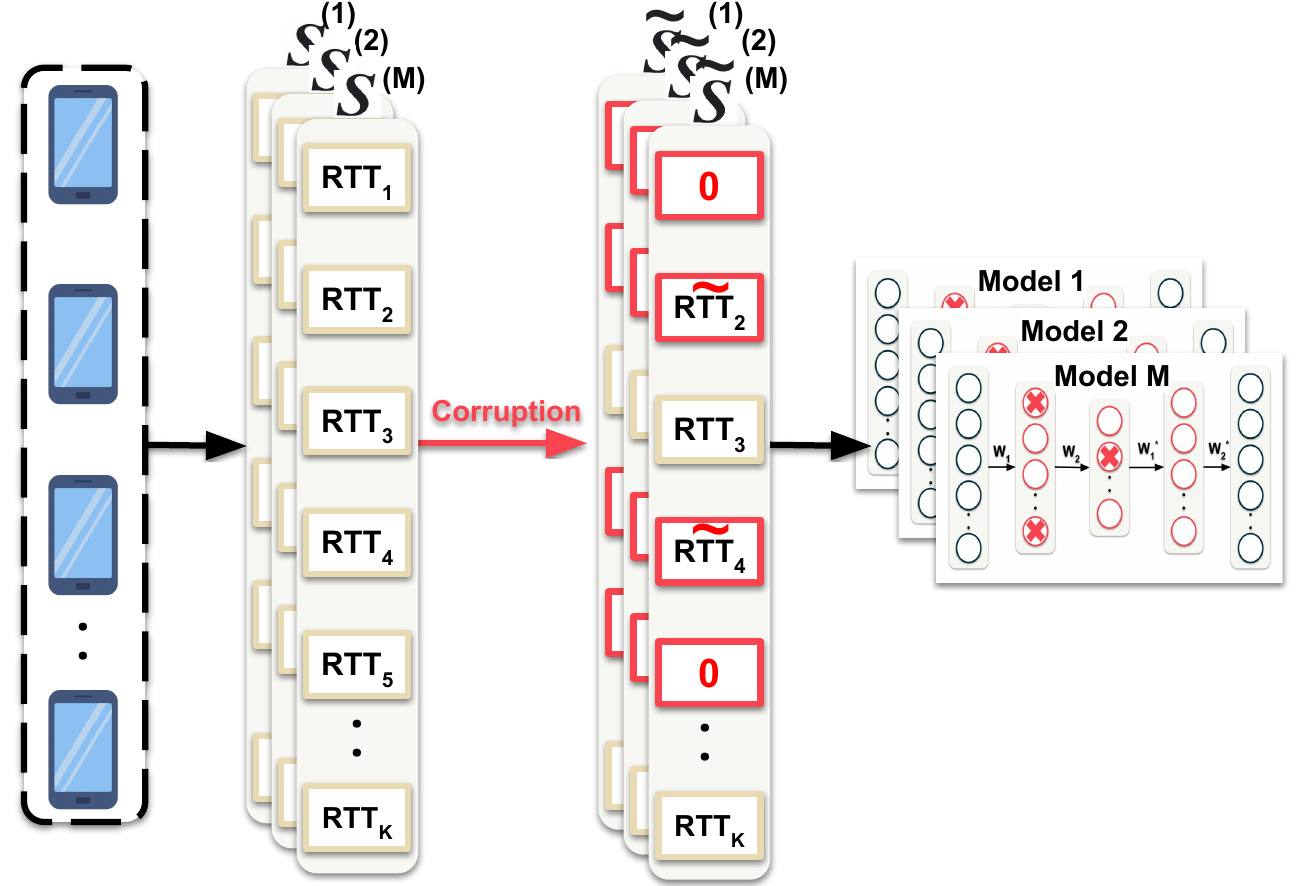}
    \caption{ \color{black}Training a stacked deep-learning auto-encoder with the RTT data of a person's existence at each reference point. Firstly, the noise is injected into the collected RTT data, then the RTT is leveraged for training. The crossed nodes are examples of dropped-out nodes during the training process. The output of each model is leveraged in a probabilistic model to estimate the object's location.}
    \label{fig:WiDAD_SysModel}  
 \end{minipage}
\end{figure}

We employ stacked denoising auto-encoders as our deep-learning models of choice, primarily due to their remarkable capability to acquire feature representations from noisy data.
Auto-encoders, in general, represent self-supervised deep-learning models that possess the inherent ability to extract latent features of lower dimensionality, which can then be used to regenerate the input data at the output \cite{vincent2010stacked}.
Denoising auto-encoders constitute an advancement over traditional auto-encoders, specifically designed to effectively handle input data corrupted by noise, such as the RTT measurements. This enhancement is achieved by deliberately introducing noise to the input data prior to its ingestion into the auto-encoder. Consequently, this drives the hidden layers of the auto-encoder to learn latent features of greater importance, thereby bolstering the model's robustness (refer to Figure~\ref{fig:WiDAD_SysModel}).
Specifically, the denoising auto-encoder is trained by initially corrupting the input state vector $s$, representing the RTT, resulting in the vector $\tilde{s}$. Our objective is to learn the parameters of the hidden layer $\mathnormal{h}$ in a manner that ensures the output of the auto-encoder $\hat{s}$  matches the uncorrupted input vector $s$. By utilizing a noisy version of the input data, the auto-encoder is compelled to capture the latent features intrinsic to the input data. It is worth noting that the weights connecting the hidden and output layers are the transpose of the weights between the input and hidden layers, reflecting the decoding process inherent to the auto-encoder.

For the rest of this section, we shall discuss how the noise is applied to the input data and how the model is trained.
\begin{figure}[t]
 \centering
 \begin{minipage}[b]{0.94\linewidth}
\includegraphics[width=0.9\linewidth,height=4.0cm,]{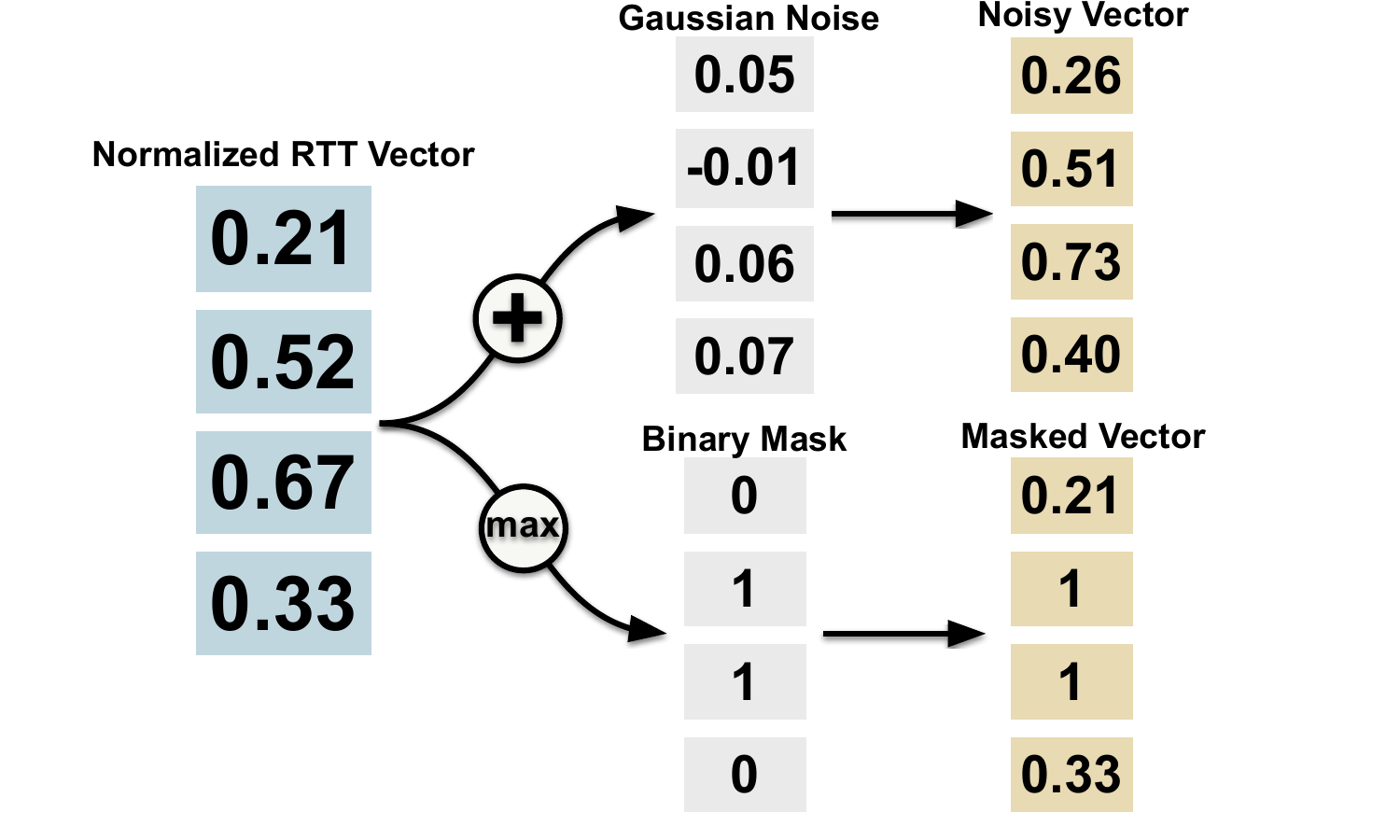}
    \caption{Examples of different corruption techniques applied to the normalized RTT states.}
    \label{fig:Curruption}  
 \end{minipage}
\end{figure}

\subsubsection{Noise Synthesizer}
In this phase, the module corrupts the input data by applying noise to the collected training samples. This increases the robustness of the system to handle noisy input data and reduces over-fitting leading to enhanced model generalization. The input corruption task is achieved by randomly adding noise to it through two techniques:  Masking corruption and Noise corruption.

\textbf{Masking corruption:}   This technique is employed based on the observation that when a Wi-Fi device, such as an Android smartphone, continuously collects RTT readings, it may not successfully detect all APs present in the surrounding area due to factors like multi-path and fading effects \cite{youssef2003small}. In order to enable the model to learn and reconstruct the original vector even with this limitation, we deliberately corrupt the original version by generating a new sample representing non-detected APs.
To accomplish this, a binary vector of size $K$ is generated, consisting of elements that are either 0 or 1. The probability of an element being one is determined by the "silence corruption factor" denoted as $P_{silence}$. The ones entries in the binary vector are then placed into the corresponding entries in the input vector,
resulting in a noisy vector where one entries indicate the non-detected APs (see Figure~\ref{fig:Curruption}).
Training the denoising autoencoder with these corrupted samples helps the model reconstruct the original samples. By learning to recover the missing AP information, the model gains the ability to handle the incompleteness caused by non-detected APs in real-world scenarios. This facilitates better generalization and performance when the model is deployed in environments where certain APs may not be consistently detected.

\textbf{Noise corruption:} The utilization of white Gaussian corruption in this technique is based on the observation that certain returned RTT values are zero, despite the actual distance between the AP and the phone never being zero. Furthermore, it has been noted that some of the reported estimated ranges even exhibit negative values. These occurrences can be attributed to the lack of calibration of the phones, which introduces an offset in the estimated range. This behavior is depicted in Figure \ref{fig:disturbance}. Additionally, some of the reported RTT-based ranges exhibit unrealistically large values, which may be a consequence of excessive processing delays within the operating system or on the card.
To mimic this behavior, a white Gaussian noise vector with a standard deviation $\sigma_{Gauss}$ is generated. This noise vector has a size of $K$ and is added to the input state vector, as illustrated in Figure~\ref{fig:Curruption}. 
By introducing this corruption and allowing the model to learn to reconstruct the original samples despite the presence of such noise and anomalies, the denoising autoencoder enhances its ability to handle real-world scenarios and improves its generalization capabilities.

\subsubsection{Model Training}

\sys{} adopts a deep-learning approach for training its models, utilizing a stacked denoising auto-encoder architecture. This architecture consists of $M$ individual models, where each model corresponds to a specific reference point (refer to Figure~\ref{fig:WiDAD_SysModel}). During the training phase, the models capture the latent features of the RTT inputs collected while an individual is stationed at the respective reference point, particularly in scenarios where NLoS conditions occur.
The purpose of capturing the latent features of the training inputs is to identify patterns and regularities that aid in detecting NLoS cases in the online behavior of the inputs, ultimately enabling accurate location estimation. The training process involves forwarding the training data through the models, followed by leveraging the reconstruction loss between the inputs and the reconstructed outputs. This reconstruction loss serves as a measure of dissimilarity between the original inputs and their reconstructed counterparts.

To optimize the models during training, a Gradient Descent optimizer is employed. This optimizer utilizes the reconstruction loss to adjust the weights across the various layers of each model, iteratively updating them to minimize the discrepancy between the inputs and the reconstructed outputs. The Mean Squared Error (MSE) loss function is utilized to quantify the reconstruction loss. Specifically, the MSE loss measures the average squared difference between the inputs and their corresponding reconstructed outputs. Consequently, each denoising auto-encoder model $m_i$ learns to effectively reconstruct the data associated with the reference location $l_i$ it represents.
On the other hand, the other denoising auto-encoder models, despite their ability to handle noise and reconstruct inputs from their respective training locations, encounter limitations in accurately reconstructing input samples from different locations, such as $l_i$. These limitations arise \textbf{due to variations in noise patterns, characteristics, and spatial variations across locations.}
By analyzing the discrepancies in the quality of reconstruction among the models, we exploit the variations in their abilities to identify NLoS cases. This exploitation enables \sys{} to estimate the persons' locations with greater accuracy and reliability, as elaborated in Section~\ref{sec:online}.

To mitigate the risk of overfitting, \sys{} implements dropout regularization during the training of its models. Dropout regularization is a widely used technique in deep learning that helps enhance the generalization capabilities of models. It achieves this by randomly setting a fraction of the activations to zero during each training iteration.
During the training process, dropout regularization introduces a form of noise or randomness into the model's hidden layers. This noise prevents individual neurons from relying too heavily on the presence of specific input features or co-adapting with other neurons. By randomly deactivating a fraction of neurons at each training step, dropout effectively forces the model to learn redundant representations and prevents overreliance on specific features or neuron dependencies. This encourages the model to distribute its learning across multiple neurons and prevents the emergence of excessively complex and specialized representations. 
In addition, \sys{} incorporates Early stopping as part of its training procedure to prevent overfitting and improve generalization in the model. It involves monitoring the model's performance on a validation set during training and halting the training process when the model's performance starts to deteriorate.
By incorporating early stopping, \sys{} ensures that the training process is effectively regulated and terminates at an optimal point, striking the right balance between model complexity and generalization. 

\color{black}

\begin{figure}[!t]
 \vspace{-0.5cm}
 \centering
 \begin{minipage}[b]{0.94\linewidth}
\includegraphics[width=\linewidth,height=7.0cm,]{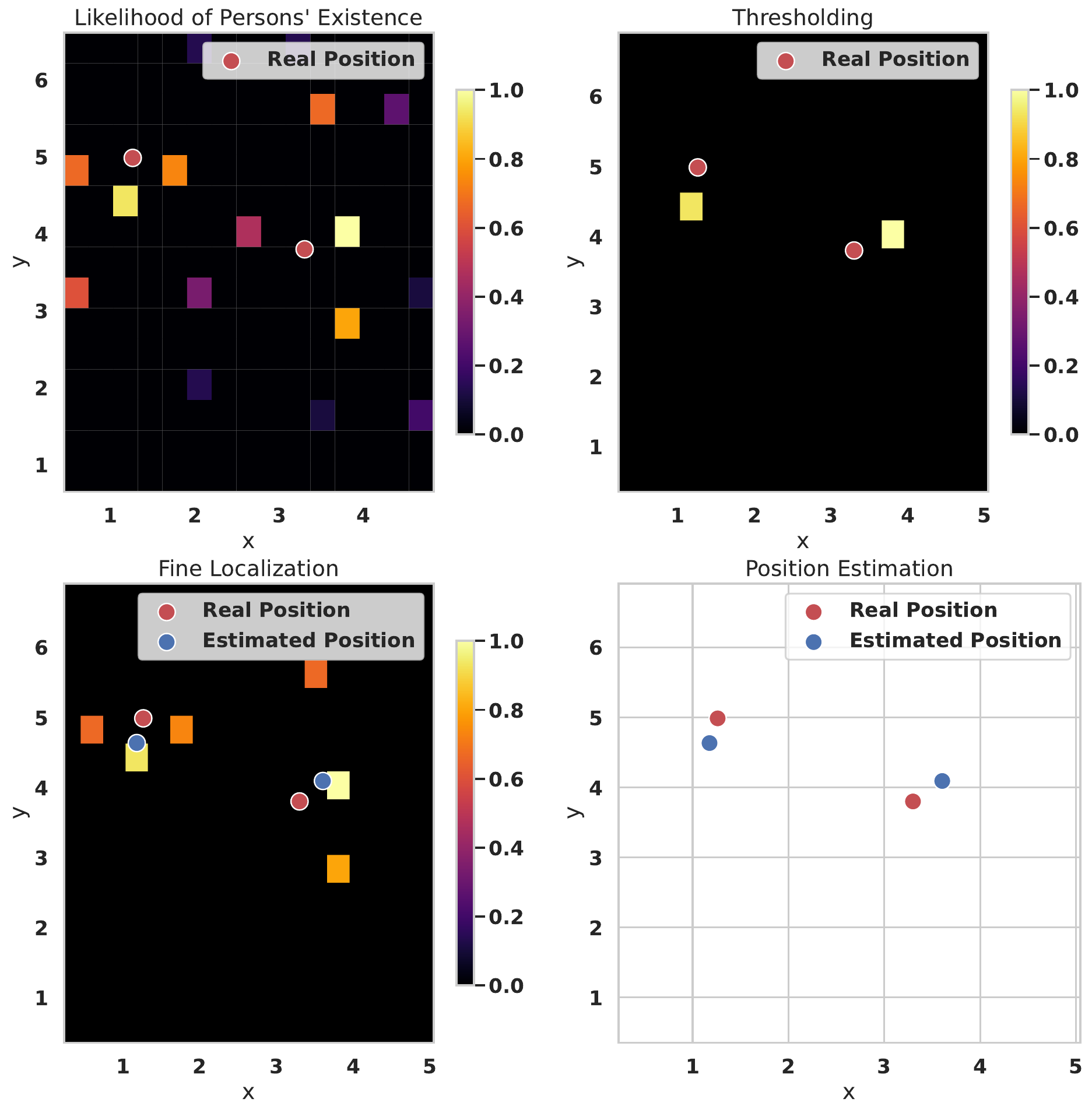}
    \caption{Multi-person localization process. First, the system normalizes the vector of the likelihood of the person's existence at each reference point. Then, it applies the threshold function to detect multi-person existence at some of the reference points. After that, the fine localization step is employed to locate the persons in the continuous space. Finally, the persons' locations are estimated}
    \label{fig:persons_2_loc}  
 \end{minipage}
  \vspace{-0.5cm}
\end{figure}

\subsection{The Online Localizer Module} \label{sec:online}

The \textbf{Online Localizer} module leverages the trained
deep-learning models to detect the persons’ existence and estimate their location. The recorded state of the system is fed to each of the trained deep-learning models. Then, the reconstructed output is leveraged in a probabilistic framework to estimate the most probable locations. The lower the reconstruction error of a model, the closer a person is to the reference point represented by that model.

Specifically, we need to find the probability of a person being at some pre-defined location $l_i$ in the area of interest given the current state vector $s_t$ of the system. More formally, we need to find $P(l_{i}\mid s_t)$. Recalling Bayes theorem, the posterior probability $P(l_{i}\mid s_t)$ is determined by:
\begin{equation} \label{p_li_s}
    P\left(l_i \mid s_t\right)=\frac{P\left(s_t \mid l_i\right) P\left(l_i\right)}{P(s_t)}=\frac{P\left(s_t \mid l_i\right) P\left(l_i\right)}{\sum_{i=1}^M P\left(s_t \mid l_i\right) P\left(l_i\right)}
\end{equation} where $P(l_i)$ is the prior probability that the pedestrian is located at a given location $l_i$ and $M$ is the number of reference points. Assuming that the system lacks information on the motion profiles of individuals, thereby treating all locations as equally probable, Equation (\ref{p_li_s}) can be reformulated as follows:
\begin{equation} \label{p_li_s_2}
    P\left(l_i \mid s_t\right) = \frac{P\left(s_t \mid l_i\right)}{\sum_{i=1}^M P\left(s_t \mid l_i\right)}
\end{equation}

In order to calculate $P\left(s_t \mid l_i\right)$, \sys{} measures the similarity score between the input state sample and each reconstructed state sample produced from each trained deep-learning model. In order to obtain the similarity score, we use a radial basis kernel (RBF) as a similarity function and since its output is between 0 and 1, it can be interpreted as the probability $P\left(s_t \mid l_i\right)$ for the $i^{th}$ model as follows:
\begin{equation} \label{p_s_li}
    P\left(s \mid l_i\right)=\frac{1}{n} \sum_{j=1}^n e^{-\frac{\left\|s_{i j}-\hat{s}_{i j}\right\|}{\sigma}}
\end{equation}
where $s_{ij}$ and $\hat{s}_{ij}$ are the original and the reconstructed input states of the $j^{th}$ scan respectively, $\sigma$ is the variance of the input scans, 
and $n$ is the total number of scans used for location estimation. 

Assuming the number of persons within the environment is designated as $N$, the system employs a multi-person localization strategy. This process begins with the normalization of the likelihood vector, representing the probability of a person's presence at each reference point, denoted as $\left\{P\left(s \mid l_i\right)\right\}_{i=1}^M$, where $M$ is the total number of reference points, and $l_i$ represents the $i^{t h}$ location. The mathematical formalization of the normalization process can be expressed as follows:
\begin{equation}
P_{\text{norm}}(s \mid l_i) = \frac{P(s \mid l_i)}{\sum_{j=1}^M P(s \mid l_j)}, \quad \forall i \in {1, 2, \ldots, M}
\end{equation}

Subsequently, a threshold function, $\tau$, is applied to $P_{\text {norm }}\left(s \mid l_i\right)$ to identify reference points with a significant likelihood of a person's presence. This can be defined as: 
\begin{equation}
\text{Detected}(l_i) = \begin{cases}1 & \text{if } P_{\text{norm}}(s \mid l_i) > \tau \\ 0 & \text{otherwise}\end{cases}
\end{equation}

Therefore, the locations of the $N$ persons are estimated as $\left\{l_1, l_2, \ldots, l_N\right\}$. Following detection, the fine localization step is undertaken to pinpoint individuals within the continuous space accurately. This step capitalizes on the spatially weighted sum of the locations of each point's nearest $K$-neighbors, multiplied by their respective normalized probabilities. This refined process is encapsulated in the equation:
\begin{equation}
L_{\text{fine}}(l_i) = \sum_{k=1}^{K} l_{k} \cdot P_{\text{norm}}(s | l_{k}), \quad l_{k} \in \text{KNN}(l_i)
\end{equation}
Here, $L_{\text {fine }}\left(l_i\right)$ represents the enhanced position for the reference point $l_i$, achieved by aggregating the products of the nearest neighbors' locations $l_k$ and their corresponding normalized probabilities. This approach allows for a more accurate estimation of each individual's position by leveraging the combined spatial and probabilistic insights from neighboring points (see Fig.~\ref{fig:persons_2_loc}). By adopting this strategy, the system not only enhances its precision in determining the positions of individuals within the environment in continuous space rather than discrete reference points but also adapts to the complexities introduced by varying densities and distributions of individuals, as verified in~\cite{youssef2004continuous}.

\section{Evaluation} \label{Evaluation}
In this section, we evaluate \sys{} within two real indoor environments, referred to as Testbed1 and Testbed2. The details of these environments are summarized in Table \ref{table:testbed}. We begin by describing the data collection process and setup. Subsequently, we conduct an ablation study to evaluate the system's performance by varying the different system parameters. Finally, we compare \sys{}'s performance with well-known existing device-free systems.

\subsection{Experimental Setup and Data Collection}

\begin{figure}[!t]
 \centering
 \begin{minipage}[b]{0.94\linewidth}
\includegraphics[width=\linewidth]{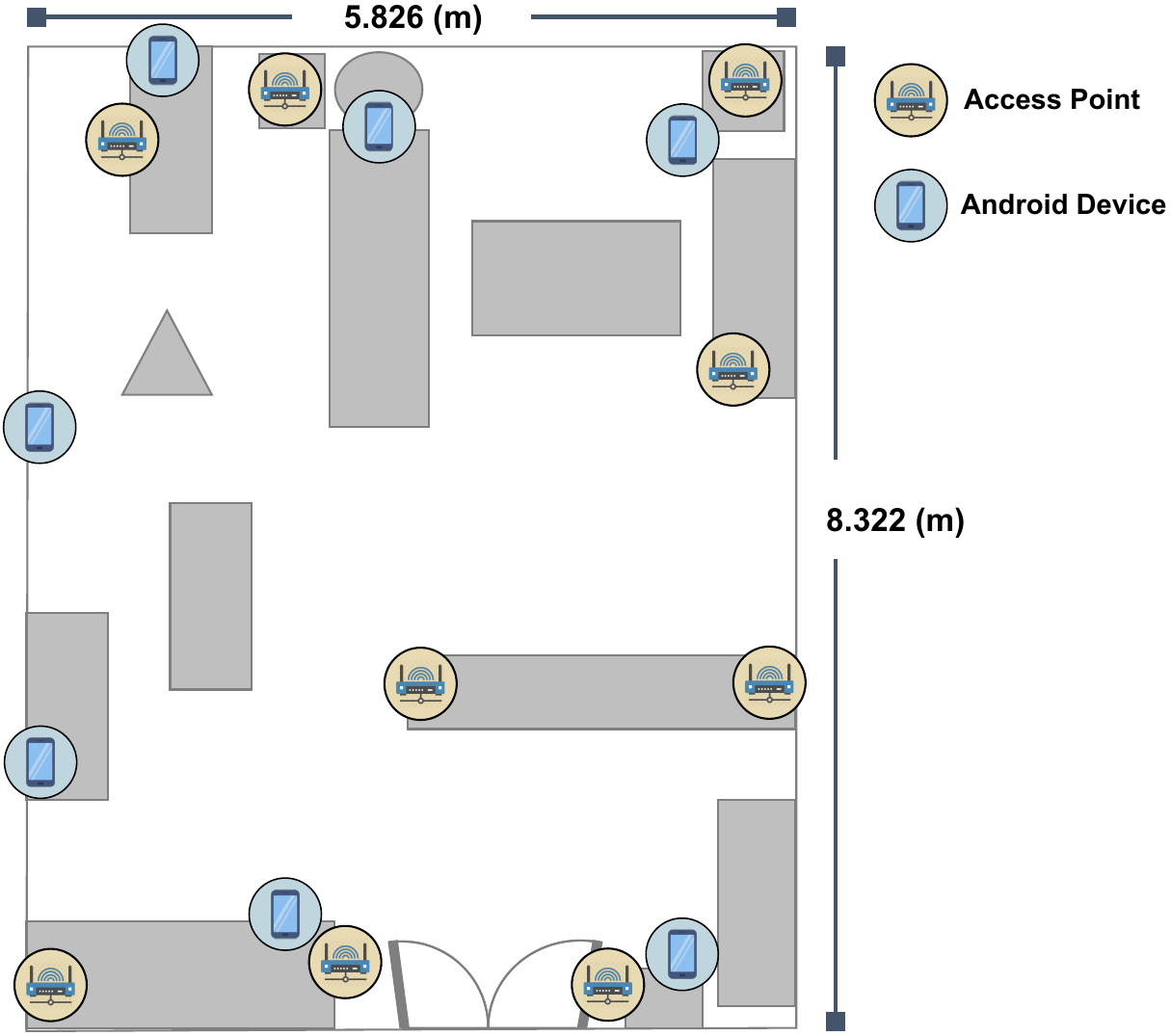}
    \caption{Testbed1.}
    \label{fig:testbed}  
 \end{minipage}
\end{figure}

\begin{figure}[!t]
 \centering
 \begin{minipage}[b]{0.94\linewidth}
\includegraphics[width=\linewidth]{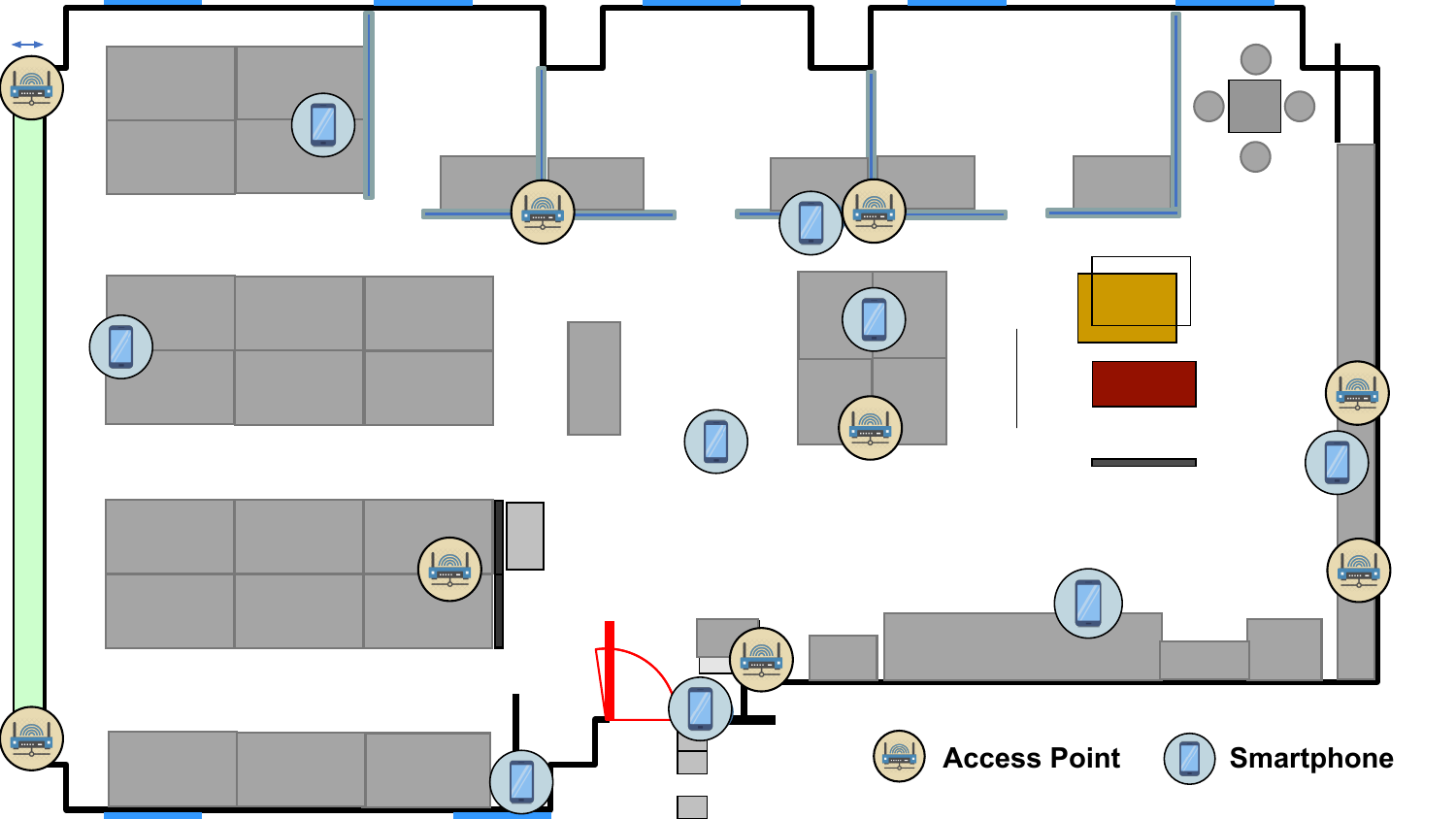}
    \caption{Testbed2.}
    \label{fig:testbed2}  
 \end{minipage}
\end{figure}

{
\begin{table}[!t]
        \centering
        \caption{Testbed Summary.}
        \label{table:testbed}
        \begin{tabular}{|c|c|c|}
        \hline
        \textbf{Testbed Parameters} & \textbf{Testbed1} & \textbf{Testbed2} \\ \hline \hline
        \textit{Area} & $5.8m \times 8.3m$ & $17.3m \times 10.9m$ \\ \hline
        \textit{Number of Access Points} & 9 & 9 \\ \hline
        \textit{Number of Android Devices} & 7 & 9 \\ \hline
        \textit{Number of Training Points} & 14 & 14 \\ \hline
        \textit{Number of Testing Points} & 4 & 10 \\ \hline

        \end{tabular}
\end{table}
}

{
\begin{table}[!t]
\color{black}
            \centering
            \caption{Default Parameters Values used in The Evaluation.}
            \label{table:parameters}
            \begin{tabular}{|l|l|l|}
            \hline
            \textbf{Parameter} & \textbf{Range}  & \textbf{Default} \\ \hline \hline
            \textit{ Dropout rate (\%)}& 0-50 & 30 and 10 \\ \hline
            \textit{ Number of Patience Epochs}& 50 & 50 \\ \hline
            \textit{ Silence Corruption Factor ($P_{silence}$) }& 0-0.5 & 0.1 \\ \hline
            \textit{ Gaussian Standard Deviation ($\sigma_{Gaussian}$)}& 0-0.5 & 0.1 \\ \hline

        
            \end{tabular}

    \end{table}
}

The system was deployed in two real-world environments. The first testbed is a room measuring $5.8m \times 8.3m$, containing offices, a meeting area, and furniture. The second testbed is a crowded lab equipped with cubicles and various types of furniture, spanning an area of  $17.3m \times 10.9m$. Each area was segmented into 18 distinct reference points distributed throughout the space. The system's infrastructure utilized transmitters from Google Wi-Fi APs and Google Nest Wi-Fi APs, with receivers being Android Google Pixel 3 devices.

\begin{figure}[!t]
    \centering
        \centering        \includegraphics[width=0.9\linewidth,height=4.5cm,]{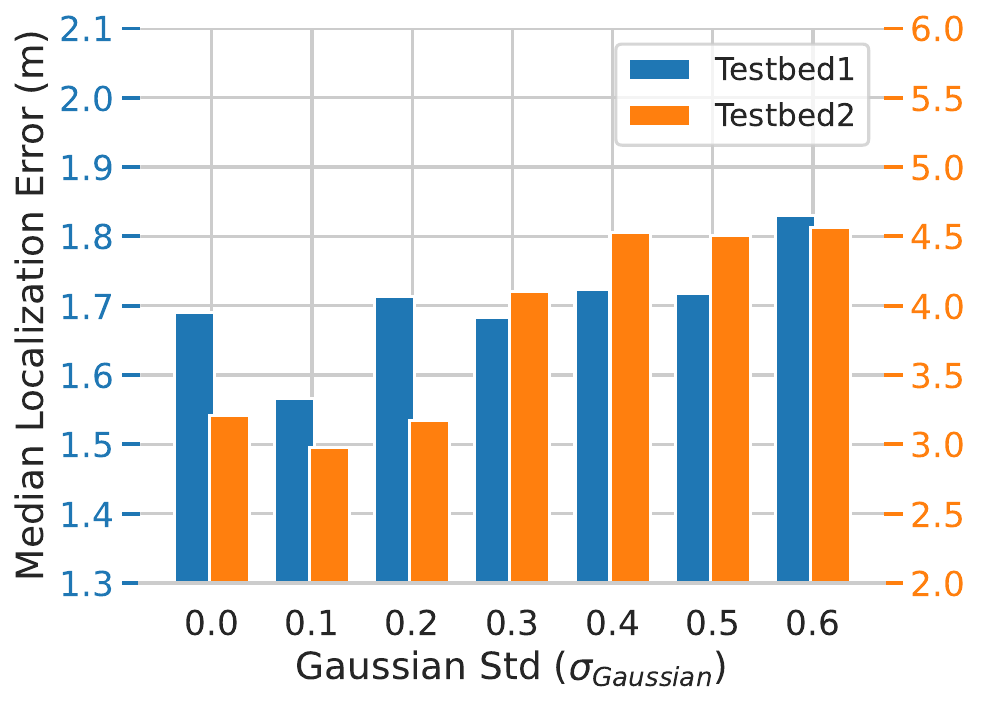}
        \caption{ Effect of the Gaussian noise standard deviation.}
        \label{fig:Gaussian_std_eval}   
    \end{figure}
   \begin{figure}[!t]
\includegraphics[width=0.9\linewidth,height=4.5cm,]{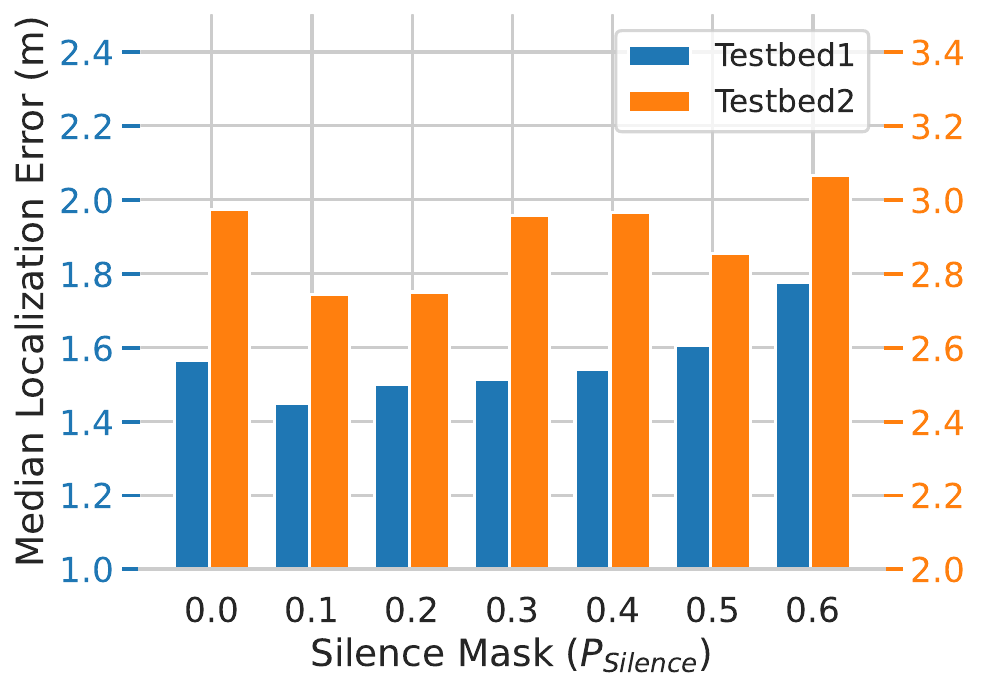}
        \caption{  Effect of the binary masking factor on accuracy.}
        \label{fig:masking_factor_eval} 
\end{figure}

\begin{figure*}[t]
    \centering
    \begin{minipage}{0.46\linewidth}
        \centering
        \includegraphics[width=\linewidth,height=4.5cm,]{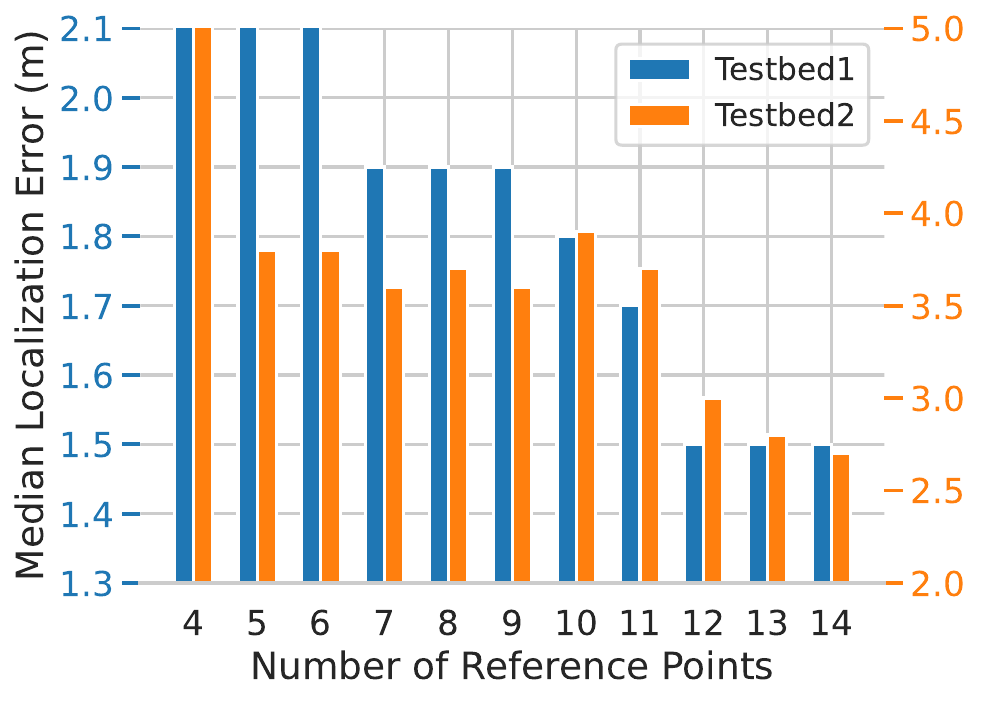}
        \caption{ Effect of reducing the number of training points.}
        \label{fig:n_training_points_eval}   
    \end{minipage}
    \begin{minipage}{0.46\linewidth}
        \centering
        \includegraphics[width=\linewidth,height=4.5cm,]{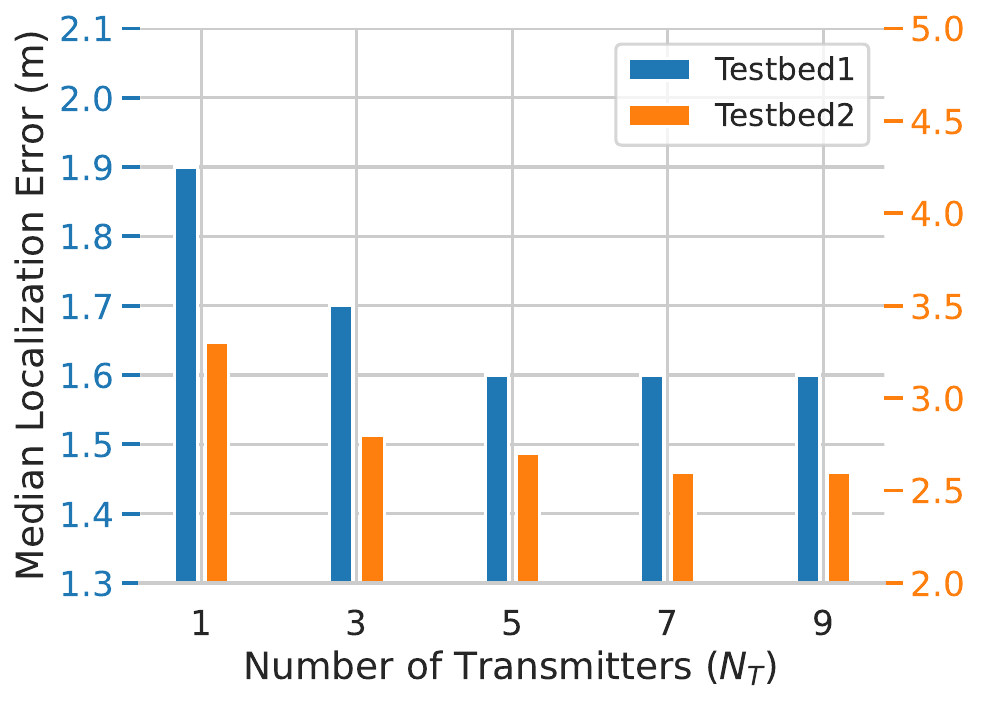}
        \caption{ Effect of reducing the number of access points.}
        \label{fig:n_APs_eval}   
    \end{minipage}
\end{figure*}

\begin{figure*}[!t]
    \centering
    \begin{minipage}{0.46\linewidth}
        \centering
        \includegraphics[width=\linewidth,height=4.5cm,]{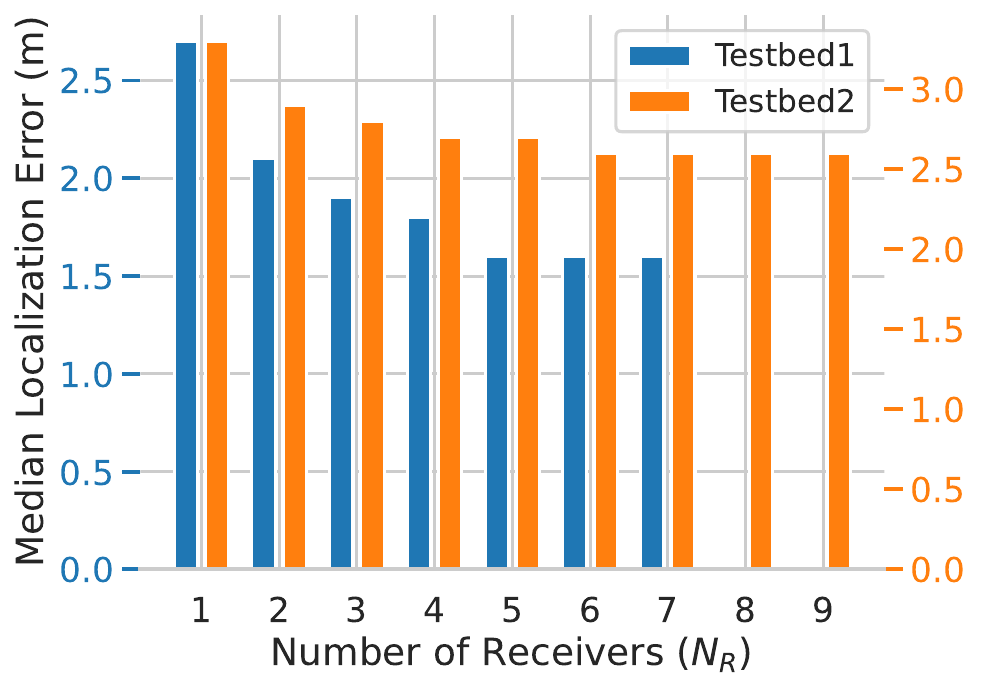}
        \caption{ Effect of reducing the number of smartphones.}
        \label{fig:n_devices_eval}   
    \end{minipage}
    \begin{minipage}{0.46\linewidth}
        \centering
        \includegraphics[width=\linewidth,height=4.5cm,]{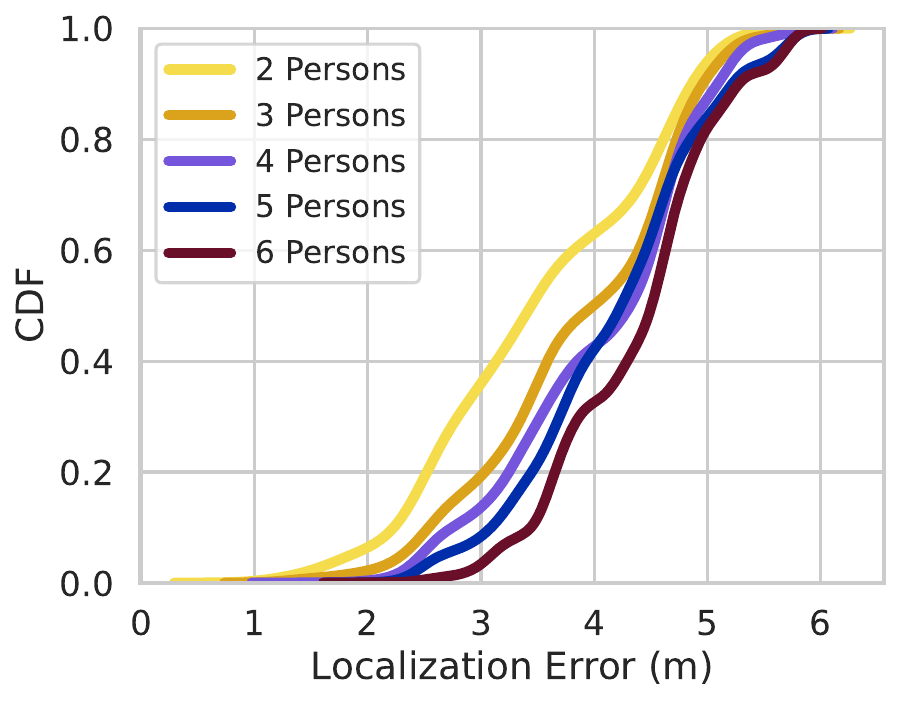}
        \caption{ CDFs of Multiple Persons.}
        \label{fig:CDF_2P}  
    \end{minipage}


\end{figure*}

Data collection was facilitated through an Android application installed on Android phones, designed to continuously scan for nearby APs. To streamline the process, the application was synchronized across all devices, with one device designated to initiate the collection. Participants could input the ground-truth coordinates of their current locations via the application interface. Additionally, data collection overhead was reduced by leveraging our previously developed system, LiPhi~\cite{liphi}, which combines crowd-sourcing with LiDAR-based automatic annotation. At each reference point, a minimum of 100 samples were collected. The data collection was conducted by six different individuals over several days during working hours, to account for the variation of indoor signals over time.

\begin{table*}[!t]
\centering
\caption{Accuracy of different systems in Testbed 1 and 2.}
\footnotesize
\label{table:widad_rtt_performance_1}
  \begin{tabular}{@{} |c|c|ccccc| @{}}
    \hline
        \textbf{Testbed} & \textbf{System} & \textbf{Average} & \textbf{25$^{th}$}  & \textbf{50$^{th}$ Percent.} & \textbf{75$^{th}$} & \textbf{Max.} \\
         &               &                  &  \textbf{Percent.} & \textbf{(Median)} & \textbf{Percent.} & \\
    \hline 
         1 & \textbf{\sys{}} & \textbf{1.61m (67\%)} & \textbf{0.85m (84\%)} & \textbf{1.57m (49\%)} & \textbf{2.25m (65\%)} & \textbf{4.87m (19\%)} \\
        & Wi-Fi RSSI \cite{youssef2007challenges} & 2.69m & 1.56m & 2.34m & 3.72m & 5.80m \\
    \hline
       2 & \textbf{\sys{}} & \textbf{3.01m (54\%)} & \textbf{2.03m (43\%)} & \textbf{2.65m (104\%)} & \textbf{3.90m (41\%)} & \textbf{6.0m (37\%)} \\
       & Wi-Fi RSSI \cite{youssef2007challenges} & 4.65m & 2.91m & 5.40m & 5.51m & 8.22m \\
    
    \hline
  \end{tabular}
\end{table*}

\subsection{Ablation Study}
{\color{black}
In this section, we study the effect of changing different system parameters on the system's performance. 
We set a maximum number of training epochs for 3000 and 50 patience epochs for the early stopping. Table \ref{table:parameters} shows the different parameters' values used throughout the evaluation section.
}

\subsubsection{Effect of input noise corruption techniques} 
In this subsection, we examine the impact of input noise corruption techniques on localization performance. Specifically, we explore the influence of two types of noise introduced during the training phase: binary masking noise and white Gaussian noise. Fig.~\ref{fig:Gaussian_std_eval} illustrates the median localization error of \sys{} when trained with varying standard deviations of Gaussian noise ($\sigma_{Gaussian}$). Similarly, Fig.~\ref{fig:masking_factor_eval} demonstrates the effect of altering the silence corruption factor ($P_{silence}$ ) on the system's median localization error.
The results from both figures indicate that incorporating noise into the training data can enhance system performance, as compared to training with noise-free inputs. This improvement is attributed to the effect of the corrupted training samples in enhancing the flexibility of the denoising autoencoders, enabling them to effectively work even with the presence of noisy, and unseen RTT measurements. However, an excessive increase in noise levels leads to a decline in system accuracy. This decline is due to the distortion of the input signal, which may introduce ambiguity in distinguishing between different locations. Optimal system performance is observed when 
$\sigma_{Gaussian} = 0.1$ and 
$P_{silence} = 0.1$, indicating a balance between noise-induced generalization and the preservation of input integrity.

\subsubsection{The Number of Training Reference Points}
{\color{black}
This section investigates the impact of varying the number of training reference points on the performance of \sys{}. For this experiment, a subset of training points was randomly selected and utilized to train the system's model. It is worth noting that reducing the number of reference points is advantageous as it decreases both the data collection overhead and the computational requirements for training and inference. Fig.~\ref{fig:n_training_points_eval} illustrates that \sys{}'s performance maintains a localization accuracy of approximately below 2m in Testbed1 and 3.5m in Testbed2 when the number of reference points is reduced to 7 and 5 respectively. This can be attributed to the system's ability to smooth the estimated locations using the probability of the neighboring points, even with a coarser grid of reference points. Although a higher density of reference points typically results in more blockage of LOS transmissions and thus better location recognition by the model, correlating RTT measurements with specific locations more precisely, our experiment demonstrates that there is an optimal number of reference points {\color{black}(7 and 5 points)} beyond which the benefits plateau. This balance between reference point density and system performance underscores the efficiency of \sys{}'s modeling and its ability to generalize from limited data. Consequently, this optimization not only preserves the system's precision but also significantly reduces the burdens associated with extensive data collection and model training.
}

\subsubsection{The Effect of the Number of Transmitter/Receiver Devices}
This experiment investigates the impact of changing the number of transmitters (access points, $N_T$) and receivers (Android devices, $N_R$) on the performance of \sys{}. Figures \ref{fig:n_APs_eval} and \ref{fig:n_devices_eval} show that \sys{} consistently achieves a localization error of approximately 1.5m in Testbed1 and 2.5m in Testbed2, even when the system operates with nearly 50\% (5 APs or 5 Android devices) of the initially deployed transmitters or receivers. This outcome, despite the reduction in LOS connections—which are pivotal for precise user localization—underscores the system's robustness under constrained conditions. This consistent performance can be attributed to two pivotal factors: Model Resilience and Location Estimation Refinement. The former is achieved through the model's training procedure, which incorporates corrupted samples. Consequently, this boosts the model's flexibility in counteracting the diminished LOS connectivity. In other words, the model is preconditioned to handle and accurately interpret suboptimal signal environments, thereby compensating for the reduced hardware setup. The latter factor is realized in the online phase, where \sys{} employs a probabilistic fine-tuning approach to the estimated user locations. This refinement process effectively mitigates potential inaccuracies that might arise from limited LOS, smoothing the localization results and enhancing reliability. 
Notwithstanding these compensatory mechanisms, a critical threshold exists beyond which a further decrease in the number of transmitters or receivers adversely affects the system's accuracy. The reason for this degradation is the exacerbated lack of LOS measurements, which introduces blind spots and amplifies ambiguity in the localization process. Thus, while \sys{} demonstrates a commendable capacity to maintain operational effectiveness with fewer devices, the experiment underscores the importance of optimizing device deployment to ensure coverage and minimize localization errors, especially in complex environments.

\begin{figure}[!t]
    \centering
        \centering
        \includegraphics[width=0.9\linewidth,height=4.5cm,]{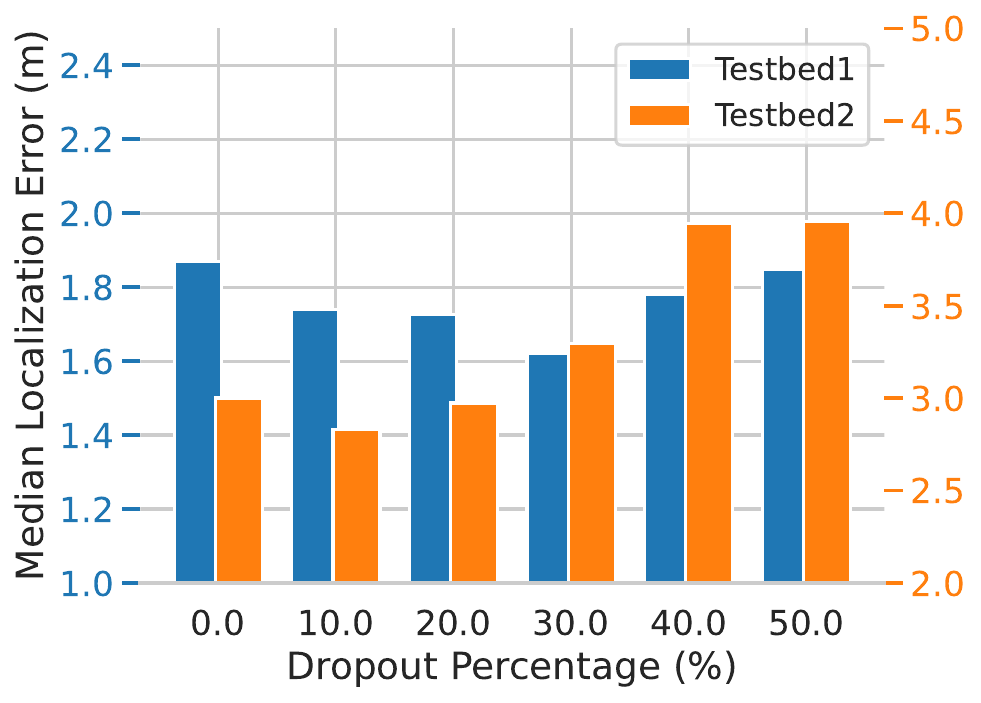}
        \caption{ The effect of changing the dropout percentage on the localization accuracy.}
        \label{fig:dropout}  
\end{figure}

{\color{black}
\subsubsection{Dropout Percentage}
The effect of changing the dropout percentage is shown in Fig.~\ref{fig:dropout}. It is obvious from the figure that at a dropout rate of 30\% and 10\%, the best performance of \sys{} is achieved on testbed1 and testbed2 respectively. This confirms the role of dropout regularization in boosting the generalizability of the trained model and ensures its resilience to over-fitting the training data. 
However, the model tends to under-fit the training data at larger dropout rates as many neurons are dropped off. 
}

\begin{figure*}[!t]
    \centering
        \centering
        \includegraphics[width=\linewidth,height=4.5cm,]{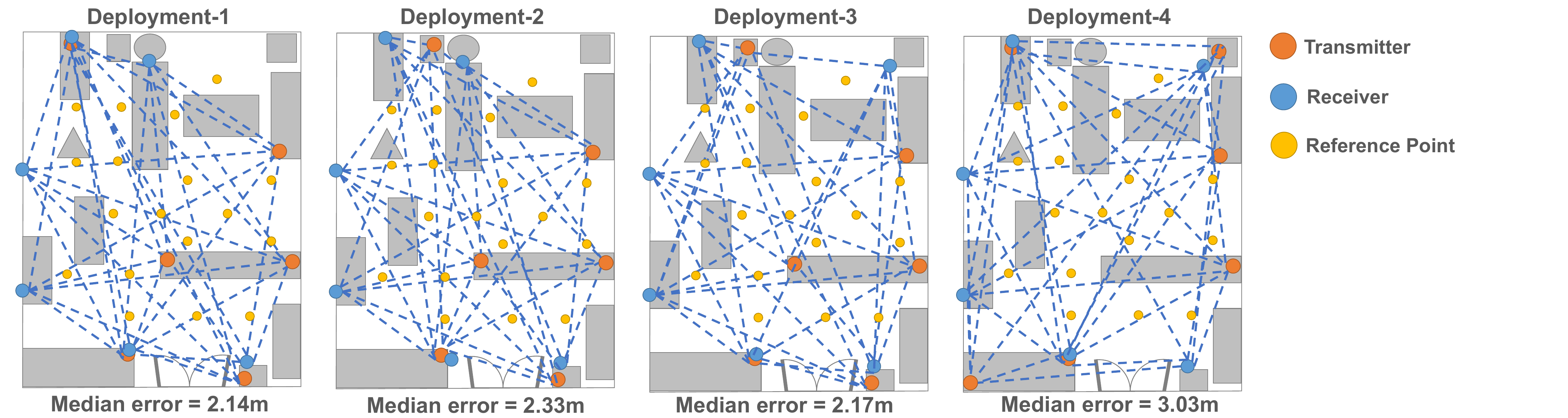}
        \caption{Different deployments' effect on accuracy.}
        \label{fig:TimeSense-Deployments}  
\end{figure*}

\begin{figure}[!t]
    \centering
        \centering
        \includegraphics[width=0.9\linewidth,height=4.5cm,]{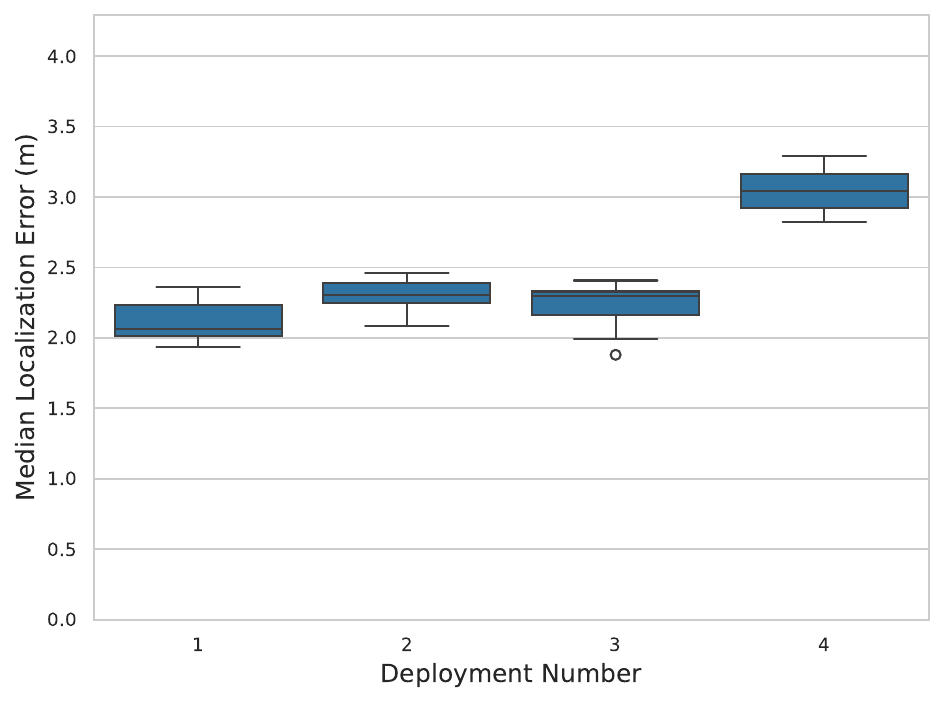}
        \caption{ Observed Localization error in each Deployment.}
        \label{fig:Deployment_boxplot}  
\end{figure}

{\color{black}
\subsubsection{The effect of changing the system's deployment}
To evaluate the effect of deployment configurations on localization accuracy, we conducted a series of experiments utilizing six transmitters and six receivers for each deployment scenario. The outcomes of these experiments are depicted in Fig.~\ref{fig:TimeSense-Deployments}\&~\ref{fig:Deployment_boxplot}. The results indicate that the system maintains consistent performance when the reference points within the test area are adequately covered, as observed in Deployments 1, 2, and 3. However, in Deployment 4, the system exhibited a decline in accuracy, attributable to the insufficient coverage of the area. These findings highlight the critical importance of ensuring comprehensive coverage of reference points through careful and strategic placement of transmitters and receivers to optimize localization accuracy.
}
\begin{figure}[!t]
    \centering
        \centering
        \includegraphics[width=0.9\linewidth,height=4.5cm,]{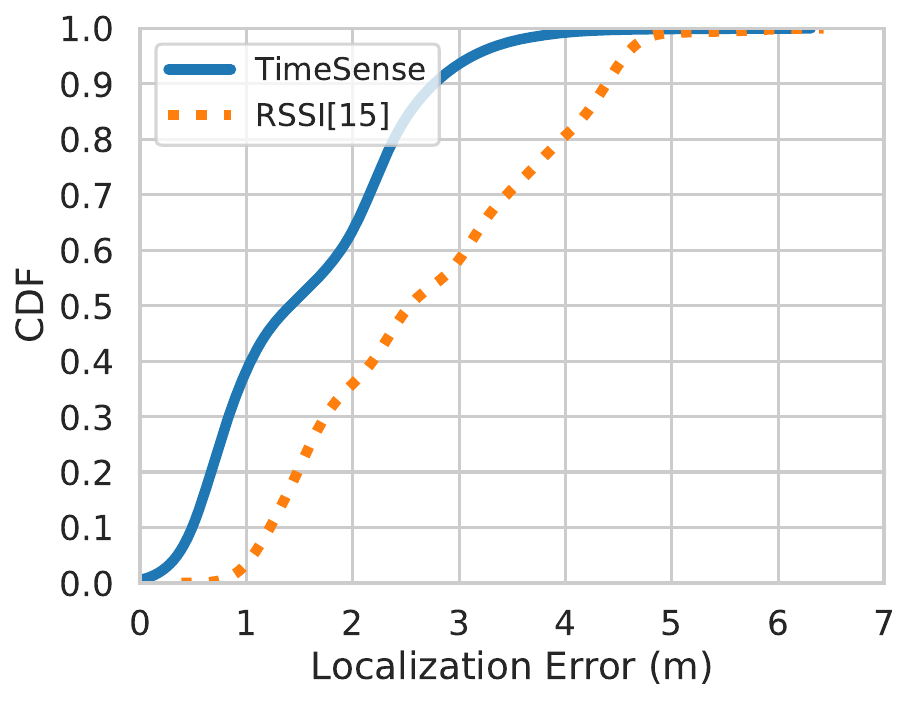}
        \caption{ CDFs of different systems' models in Testbed 1.}
        \label{fig:CDF1}  
\end{figure}

\subsubsection{The Effect of the Number of Persons in the Room}
{\color{black}
In this experiment, we evaluate \sys{}'s localization performance when testbed1 is populated with multiple individuals, focusing on how increased human presence impacts the system. It's widely recognized that an increased number of people in a room typically leads to a notable decline in the efficacy of radio-based localization systems. This decline is attributed to interference in radio measurements (e.g., RSSI), which introduces additional noise, amplifies multipath effects, and exacerbates signal reflections and occlusions. These factors collectively complicate the interpretation of signals and increase the probability of localization inaccuracies.

Contrastingly, the proposed RTT-based system, \sys{}, exhibits only a modest reduction in performance—quantified at the decimeter level—with the addition of each person, as depicted in Fig.~\ref{fig:CDF_2P} 
These findings demonstrate the system's localization precision with up to six individuals spread across the environment. This minimal performance degradation, coupled with remarkable resilience, is underpinned by the inherent robustness of RTT measurements, which confirm a notable advantage over traditional technologies like RSSI \cite{feng2022analysis}. Additionally, \sys{} benefits from advanced denoising techniques and a probabilistic fine-tuning method. These strategic implementations empower \sys{} to adeptly navigate the increased noise and complexity introduced by a denser human presence, affirming the system's capacity to sustain accuracy amidst rising environmental density.
}

\begin{figure}[!t]
    \centering
        \centering
        \includegraphics[width=0.9\linewidth,height=4.5cm,]{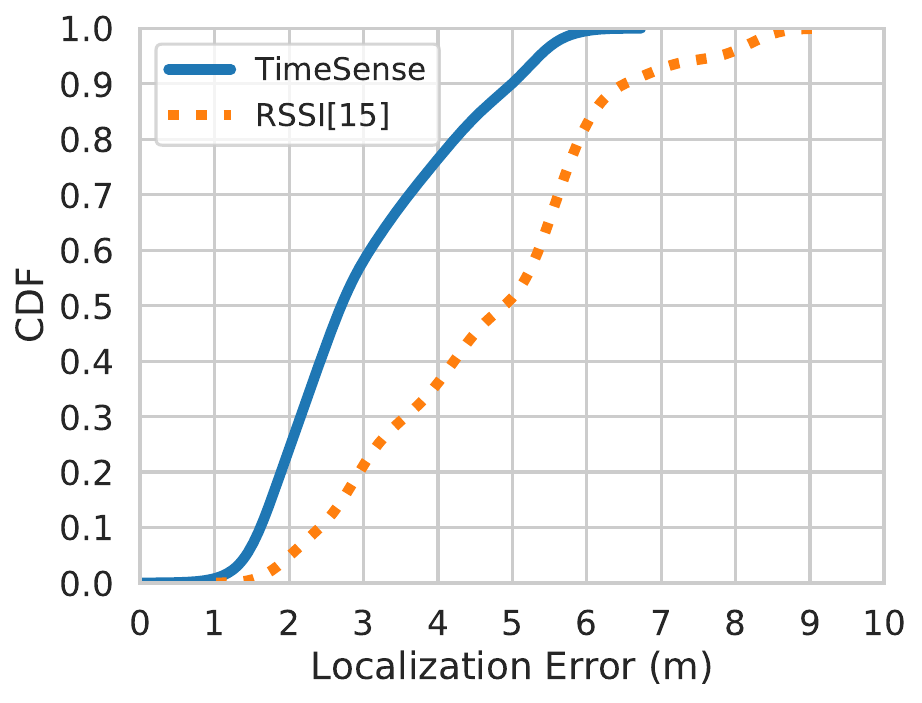}
        \caption{ CDFs of different systems' models in Testbed 2.}
        \label{fig:CDF2}  
\end{figure}

\subsection{Comparative Evaluation}
In this section, we present a comparative analysis of \sys{}'s localization performance in relation to a state-of-the-art system proposed in \cite{youssef2007challenges} that utilizes Wi-Fi received signal strength indication (RSSI) for device-free localization. The RSSI-based system constructs a radio map within the target area by employing Wi-Fi RSSI measurements. During disturbances caused by the movement of individuals, a probabilistic model is utilized to accurately localize by comparing the collected RSSI data with pre-collected data during the offline phase. We have evaluated our system's model using RSSI measurements.

The evaluation results of \sys{} in two different testbeds are illustrated in Fig.~\ref{fig:CDF1} and Fig.~\ref{fig:CDF2}, and further summarized in Table~\ref{table:widad_rtt_performance_1}. As depicted in the figures and table, \sys{} demonstrates superior performance compared to the Wi-Fi RSSI system \cite{youssef2007challenges}, achieving a median localization error reduction of 49\% and 103\% in Testbed 1 and Testbed 2, respectively. This notable improvement can be attributed to several challenges faced by the RSSI-based system \cite{youssef2007challenges}, including multipath effects, interference, and variations in transmitted signal power. Moreover, the probabilistic techniques employed by the RSSI-based system, although more effective than deterministic techniques in handling the inherently noisy wireless signals, often assume independence among signals from different access points (APs) to avoid the curse of dimensionality problem \cite{nasrabadi2007pattern}. Such an assumption adversely affects its accuracy.
In contrast, \sys{} utilizes a deep neural network that effectively learns dependencies between round-trip time (RTT) measurements from different APs. This architectural choice enables \sys{} to overcome the limitations associated with the independence assumption and mitigate the inherent delays or distortions in the measurements. Furthermore, \sys{} incorporates mechanisms to address overfitting, resulting in improved robustness.

It is important to note that the higher localization error observed in Testbed 2 can be attributed to the crowded environment and presence of furniture, which lead to increased signal reflections and consequently higher levels of noise in the measurements.

\section{Conclusion} \label{sec:Conclusion}
{\color{black}
In this work, we proposed \sys{}, an indoor multi-person device-free localization system based on the standardized Wi-Fi RTT technique. \sys{} leverages a deep-learning denoising auto-encoder model to mitigate the challenging noise in the wireless channel. It applies a probabilistic localization model to estimate the likelihood of the persons' locations at the predefined reference points. Finally, the results are refined by finding the center of mass of the places with high likelihoods weighted by their corresponding likelihoods. We evaluated our system's performance in two real environments, it achieved a median accuracy of up to 1.5m. Additionally, we tested the system's capability of multi-person localization.
In future work, we plan to deploy our system in more complex and cluttered environments to evaluate its robustness and adaptability. This will include testing in environments with various obstructions and varying levels of interference. Additionally, we aim to assess the system's performance across different devices, addressing the challenges of device heterogeneity. This involves using a range of devices with different specifications and capabilities to ensure the system's broad applicability. We will also experiment with changing the locations of transmitters and receivers to determine the impact on system performance and to optimize positioning strategies. Furthermore, we intend to investigate the fusion of Round-Trip Time (RTT) data with Received Signal Strength Indicator (RSSI) and Channel State Information (CSI). This fusion aims to enhance the accuracy and reliability of our system by leveraging the complementary strengths of these different data types. Through these comprehensive evaluations and enhancements, we aim to further improve the effectiveness and versatility of our system.
}

\bibliography{_ref}
\bibliographystyle{IEEEtran}

\begin{IEEEbiography}[{\includegraphics[width=1in,height=1.25in,clip,keepaspectratio]{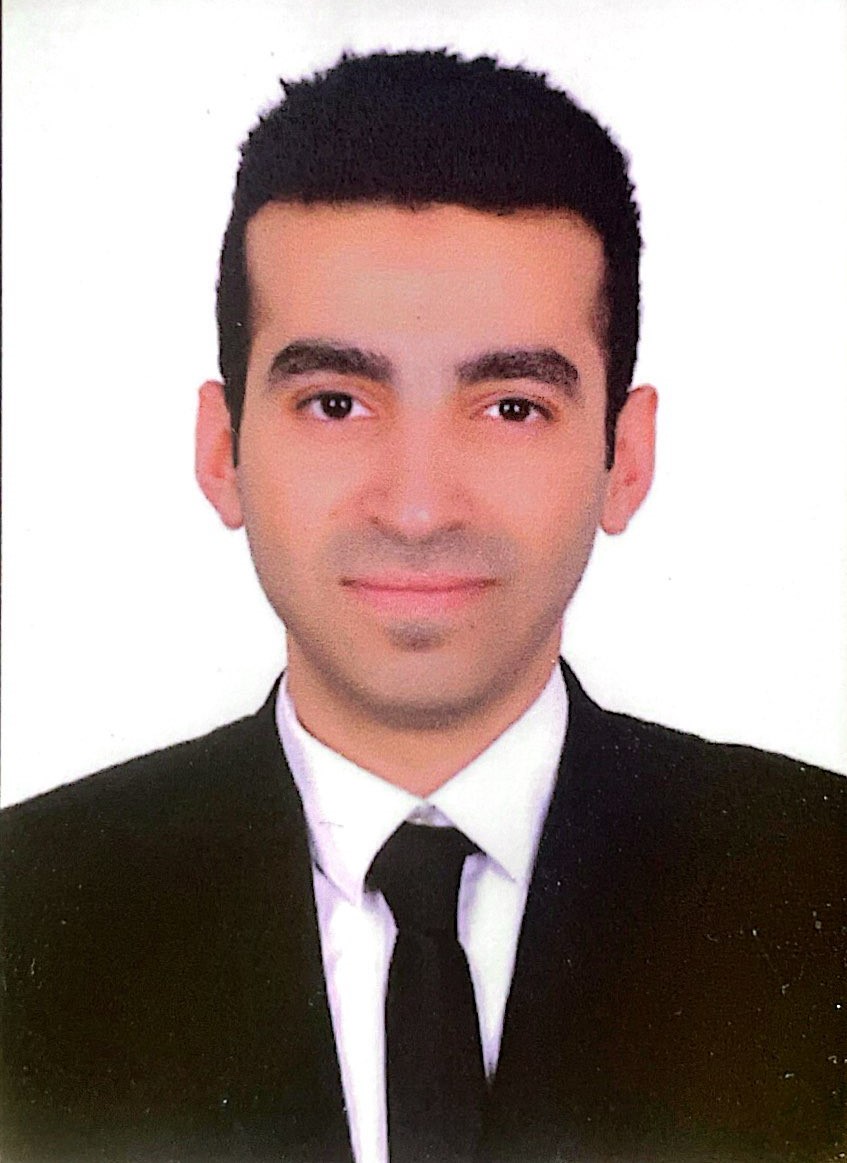}}]{Mohamed Mohsen} received the B.E.
degree in Communication and Electronics Engineering from Benha University, Egypt, in 2021. He placed $3^{rd}$ so he was offered a teaching assistant position for mathematics in the same college. He is interested in topics related to AI and Embedded Systems.
\end{IEEEbiography}

\begin{IEEEbiography}[{\includegraphics[width=1in,height=1.25in,clip,keepaspectratio]{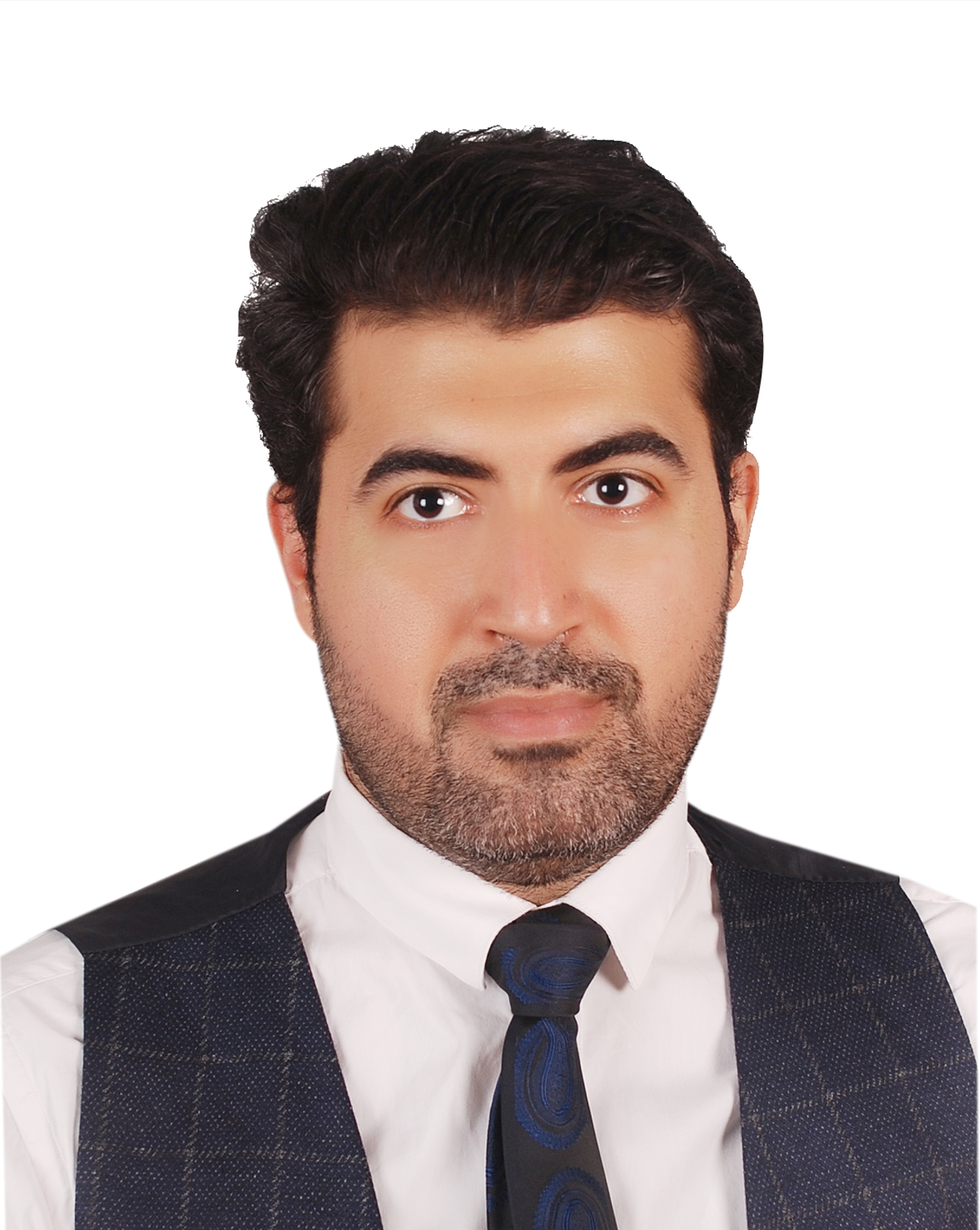}}]{Hamada Rizk} (IEEE Senior Member)
received M.E. and  Ph.D. degrees in  Computer Science and Engineering from Tanta University and E-JUST in 2016 and 2020, respectively. He is currently affiliated with Tanta University, Egypt, and Osaka University, Japan. He has been working in mobile and pervasive computing, spatial intelligence, and AI research areas. He has been involved in many projects funded by several academic and industrial organizations such as NTRA Egypt, Uber, USA, ASTEP JST, Kakenhi  JSPS, NVIDIA, Japan, etc.  He has authored several publications in top journals and conferences and holds several patents. Hamada is the recipient of the silver medal in the 4th ACM SigSpatial competition held in Chicago, in 2019. He has also been honored as an outstanding young researcher by the HLF foundation in Germany (2019) and Google (2019\&2020), among others.\end{IEEEbiography}

\vspace{-1.1cm}

\begin{IEEEbiography}[{\includegraphics[width=1in,height=1.25in,clip,keepaspectratio]{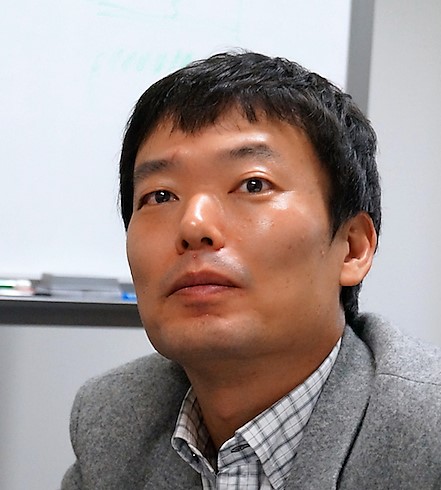}}]{Hirozmi Yamaguchi}
received the B.E., M.E., and Ph.D. degrees in information and computer science from Osaka University, Osaka, Japan in 1994, 1996 and 1998, respectively. He is currently a professor at Osaka University and leading Mobile Computing laboratory. He has been working in mobile and pervasive computing and networking research areas and has published papers in top-quality journals and conferences. He has served on the editorial board of Elsevier Ad Hoc Networks Journal and Springer Journal of Reliable Intelligent Environments. He has also served on ICDCN2021 and Mobiquitous 2021 as general chairs, and on many mobile and pervasive conferences such as IEEE PerCom as TPC chairs and members. He was awarded Commendation for Science and Technology by the Minister of Education, Culture, Sports, Science and Technology in 2018.
\vspace{-1.2cm}
\end{IEEEbiography}

\begin{IEEEbiography}[{\includegraphics[width=1in,height=1.25in,clip,keepaspectratio]{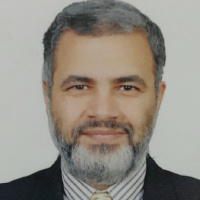}}]{Moustafa Youssef}
is a professor at the American University in Cairo and Alexandria University and founder \& director of the Wireless Research Center of Excellence, Egypt. His research interests include mobile wireless networks, mobile and pervasive computing, location determination technologies, and quantum computing. He is an Associate Editor for IEEE TMC and ACM TSAS, served as the Lead Guest Editor of the IEEE Computer Special Issue on Transformative Technologies, and an Area Editor of ACM MC2R as well as on the organizing and technical committees of numerous prestigious conferences. He is the recipient of the 2003 the University of Maryland Invention of the Year award, the 2010 TWAS-AAS-Microsoft Award for Young Scientists, the 2013 and 2014 COMESA Innovation Award, the 2013 ACM SIGSpatial GIS Conference Best Paper Award, the 2017 Egyptian State Award, multiple Google Research Awards, among many others. He is also an IEEE and ACM Fellow.
\end{IEEEbiography}

\end{document}